\documentclass[fleqn,usenatbib]{mnras}

\usepackage{newtxtext,newtxmath}

\usepackage[T1]{fontenc}

\DeclareRobustCommand{\VAN}[3]{#2}
\let\VANthebibliography\thebibliography
\def\thebibliography{\DeclareRobustCommand{\VAN}[3]{##3}\VANthebibliography}

\usepackage{graphicx}	
\usepackage{amsmath}	

\usepackage{anyfontsize}

\usepackage{xcolor}
\usepackage{xspace}

\newcommand{\fref}[1]{Fig.~\ref{#1}}
\newcommand{\tref}[1]{Table~\ref{#1}}

\newcommand{\cref}[1]{Chapter~\ref{#1}}
\newcommand{\sref}[1]{Section~\ref{#1}}
\newcommand{\aref}[1]{Appendix~\ref{#1}}

\usepackage{pdflscape}
\usepackage{afterpage}
\usepackage{caption}

\usepackage{subcaption}

\usepackage{CJKutf8}
\usepackage[T2A, T1]{fontenc}
\usepackage[russian, english]{babel} 

\usepackage[super]{nth}

\title[\textit{Kilonova Seekers}]{\textit{Kilonova Seekers}: the GOTO project for real-time citizen science in time-domain astrophysics}

\author[Killestein and Kelsey et al.]{
\parbox{\textwidth}{
T.~L.~Killestein,$^{1,2}$$^\ddagger$\thanks{E-mail: thomas.killestein@utu.fi (TLK) $^\ddagger$ Joint first authorship}
L.~Kelsey,$^{3}$$^\ddagger$\thanks{E-mail: lisa.kelsey@port.ac.uk (LK)}
E.~Wickens,$^{3}$
L.~Nuttall,$^{3}$
J.~Lyman,$^{2}$
C.~Krawczyk,$^{3}$
K. Ackley,$^{2}$
M.~J.~Dyer,$^{4}$ 
F.~Jim\'enez-Ibarra,$^{5}$ 
K.~Ulaczyk,$^{2}$ 
D.~O'Neill,$^{2}$ 
A.~Kumar,$^{2}$ 
D.~Steeghs,$^{2}$ 
D.~K. Galloway,$^{5}$
V.~S.~Dhillon,$^{4,11}$ 
P.~O'Brien,$^{6}$ 
G.~Ramsay,$^{7}$
K.~Noysena,$^{8}$ 
R.~Kotak,$^{1}$ 
R.~P.~Breton,$^{9}$ 
E.~Pall\'e,$^{11,12}$ 
D.~Pollacco,$^{2}$ 
S.~Awiphan,$^{8}$ 
S.~Belkin,$^{5}$
P.~Chote,$^{2}$
P.~Clark,$^{3}$
D.~Coppejans,$^{2}$
C.~Duffy,$^{7}$ 
R.~Eyles-Ferris,$^{6}$ 
B.~Godson,$^{2}$ 
B.~Gompertz,$^{14}$
O.~Graur,$^{3,15}$ 
P.~Irawati,$^{8}$ 
D.~Jarvis,$^{4}$
Y.~Julakanti,$^{6}$ 
M.~R. Kennedy,$^{10}$ 
H.~Kuncarayakti,$^{1}$
A.~Levan,$^{13}$
S.~Littlefair,$^{4}$
M.~Magee,$^{2}$
S.~Mandhai,$^{9}$ 
D.~Mata S\'anchez,$^{11,12}$ 
S.~Mattila,$^{1,16}$ 
J.~McCormac,$^{2}$ 
J.~Mullaney,$^{4}$
J.~Munday,$^{2}$
M.~Patel,$^{6}$ 
M.~Pursiainen,$^{2}$
J.~Rana,$^{11,12}$ 
U.~Sawangwit,$^{8}$ 
E.~Stanway,$^{2}$ 
R.~Starling,$^{6}$ 
B.~Warwick,$^{2}$ 
K.~Wiersema$^{17}$
}
\\
\\
$^{1}$ Department of Physics \& Astronomy, University of Turku, Vesilinnantie 5, Turku, FI-20014, Finland.\\
$^{2}$ Department of Physics, University of Warwick, Gibbet Hill Road, Coventry CV4 7AL, UK.\\
$^{3}$ Institute of Cosmology and Gravitation, University of Portsmouth, Portsmouth, PO1 3FX, UK.\\
$^{4}$ Department of Physics and Astronomy, University of Sheffield, Sheffield S3 7RH, UK. \\
$^{5}$ School of Physics \& Astronomy, Monash University, Clayton VIC 3800, Australia. \\
$^{6}$ School of Physics \& Astronomy, University of Leicester, University Road, Leicester LE1 7RH, UK. \\
$^{7}$ Armagh Observatory \& Planetarium, College Hill, Armagh, BT61 9DG.\\
$^{8}$ National Astronomical Research Institute of Thailand, 260 Moo 4, T. Donkaew, A. Maerim, Chiangmai, 50180 Thailand.\\
$^{9}$ Jodrell Bank Centre for Astrophysics, Department of Physics and Astronomy, The University of Manchester, Manchester M13 9PL, UK.\\
$^{10}$ School of Physics, Kane Building, University College Cork, Cork, Ireland. \\
$^{11}$ Instituto de Astrof\'{i}sica de Canarias, E-38205 La Laguna, Tenerife, Spain. \\
$^{12}$ Departamento de Astrof\'isica, Univ. de La Laguna, E-38206 La Laguna, Tenerife, Spain. \\
$^{13}$ Radboud University, Postbus 9010, 6500 GL, Nijmegen, Netherlands. \\
$^{14}$ School of Physics and Astronomy, University of Birmingham, Birmingham, B15 2TT, UK. \\
$^{15}$ Department of Astrophysics, American Museum of Natural History, Central Park West and 79th Street, New York, NY 10024-5192, USA. \\
$^{16}$ School of Sciences, European University Cyprus, Diogenes street, Engomi, 1516, Nicosia, Cyprus. \\
$^{17}$ Centre for Astrophysics Research, University of Hertfordshire, College Lane, Hatfield AL10 9AB, UK.\\
}

\date{Accepted XXX. Received YYY; in original form ZZZ}

\pubyear{2024}

\begin{document}
\label{firstpage}
\pagerange{\pageref{firstpage}--\pageref{lastpage}}
\maketitle

\begin{abstract}
Time-domain astrophysics continues to grow rapidly, with the inception of new surveys drastically increasing data volumes. Democratised, distributed approaches to training sets for machine learning classifiers are crucial to make the most of this torrent of discovery -- with citizen science approaches proving effective at meeting these requirements.
In this paper, we describe the creation of and the initial results from the \textit{Kilonova Seekers} citizen science project, built to find transient phenomena from the GOTO telescopes in near real-time. \textit{Kilonova Seekers} launched in July 2023 and received over 600,000 classifications from approximately 2,000 volunteers over the course of the LIGO-Virgo-KAGRA O4a observing run. During this time, the project has yielded 20 discoveries, generated a `gold-standard' training set of 17,682 detections for augmenting deep-learned classifiers, and measured the performance and biases of Zooniverse volunteers on real-bogus classification. This project will continue throughout the lifetime of GOTO, pushing candidates at ever-greater cadence, and directly facilitate the next-generation classification algorithms currently in development.
\end{abstract}

\begin{keywords}
techniques: miscellaneous -- (stars:) supernovae: general -- surveys
\end{keywords}


\section{Introduction} \label{Intro}
In the current era of time-domain astronomy, we are operating close to the limit of human validation of transient phenomena due to the vast numbers of observations being taken on a daily basis. The expansive data volumes (TB per night) of current all-sky surveys such as the Gravitational-wave Optical Transient Observer \citep[GOTO;][]{Steeghs2022}, Zwicky Transient Facility \citep[ZTF;][]{Bellm2019}, Asteroid Terrestrial-impact Last Alert System \citep[ATLAS;][]{Tonry2018}, and All-Sky Automated Survey for Supernovae \citep[ASAS-SN;][]{Kochanek2017}, and the impending era of the Vera C. Rubin Observatory's Legacy Survey of Space and Time \citep[LSST;][]{Ivezic2019} highlight the continuing need for novel, automated, machine-learned approaches of source classification in order to triage and follow-up candidates in a timely manner. 

Modern transient discovery is predominantly based on difference imaging (e.g. \citealt{Alard1998,Zackay2016}). In this technique, \lq{template}\rq, \lq{reference}\rq\ or \lq{background}\rq\ images are subtracted from new \lq{science}\rq\ images in order to remove non-varying sources from the image. These reference images are of the same part of the sky as the science image, but were taken at a prior time during the optimal sky conditions (dark moon phases, good seeing). Typically they are also of longer exposure than the science images, meaning that fainter sources can be detected. Subtracting the reference image from the new science image, after correcting for differential background and PSF mis-matches, results in a \lq{difference}\rq\ image. This difference image may contain residual flux indicating that something has changed between the reference and science images -- a potential transient or variable source has appeared. The photometry can then be extracted from the difference image, to measure positions and fluxes free of contamination from surrounding sources \citep[e.g.][]{Wozniak2002} or host galaxy light.

The majority of detections (referred to as candidates herein) in difference images detected via source extraction are artefacts, known as `bogus' sources following the real-bogus paradigm introduced in \citet{Bloom2012}. These artefacts broadly arise from bright star residuals, point-spread-function (PSF) mis-match, and/or misalignment. A vast literature has emerged to tackle this challenge -- transitioning from traditional machine learning (ML) approaches \citep{Bailey2007,Goldstein2015,Wright2015,Mong2020}, through to deep learned classifiers \citep{Cabrera-Vives2017,Duev2019,Killestein2021,Mong2023,Corbett2023} -- with ever increasing performance. Naturally however, as surveys grow larger, more performant source classification algorithms are required to ensure that the number of (inevitable) false positives do not overwhelm human vetters. To achieve this goal, larger and larger data volumes are required to effectively train these algorithms, and fully sample the diversity of detections seen in survey data. As surveys get bigger, the method for dealing with these data volumes needs to improve. Such surveys quickly outstrip the capacity of individuals or small teams of scientists to effectively label. A complementary approach, which can be used to create a human-labelled data-set for training machine-learning based classifiers, is to use citizen science.

Citizen science enables collaboration between researchers and members of the public, by engaging the public to participate in research tasks and help make scientific discoveries. For tasks such as vetting of candidate transients, the person-power increase of opening this task up to the public is highly significant. Transient astronomy projects on the Zooniverse citizen science platform\footnote{\url{https://www.zooniverse.org/}} such as \textit{Galaxy Zoo Supernovae} \citep{Smith2011} and \textit{Supernova Hunters} \citep{Wright_2017}, using data from the Palomar Transient Factory (PTF; \citealt{Law2009,Rau2009}) and Pan-STARRS1 \citep{Chambers2016} respectively, have had great success involving the public in this way. In both cases, volunteers were provided with a set of target, reference and difference images for a candidate transient that had been flagged as interesting by a computer algorithm, and were asked a simple question to determine if the observation was real or bogus. This facilitates discovery of transient events, and creates a binary-labelled training set for ML algorithms to augment their performance in future iterations.

Alongside the direct benefits for scientific analysis, citizen science provides an excellent opportunity for public engagement and outreach by enabling members of the public to help in key scientific discovery, and to achieve experiential learning~\citep{Bruner1961,Kolb1984}. The Zooniverse platform was originally created for the flagship \textit{Galaxy Zoo} project \citep{Lintott2008}, and has since become the predominant online platform for facilitating citizen science \citep{Marshall2015}. At the time of writing, the Zooniverse platform has 91 active projects on offer, with topics ranging from history, language, and literature to climate, nature, physics, and space; meaning that there is something of interest for everyone.
Citizen science approaches have led to tangible scientific discoveries: In astronomy, the \textit{Galaxy Zoo} project led to the discovery of `green pea' galaxies, a new class of compact, star-forming galaxies~\citep{Cardamone2009}. Similarly, the \textit{Planet Hunters} project enabled the discovery of PH1b, the first known planet in a quadruple star system~\citep{Schwamb2013}.

We have developed the \textit{Kilonova Seekers} citizen science project\footnote{\url{http://kilonova-seekers.org/}} on the Zooniverse platform, providing an opportunity for members of the public to help the GOTO collaboration in the discovery of transient events that may have been otherwise missed or overlooked, and enabling them to participate in cutting-edge science in near real-time.

In this paper, we report findings from the launch of \textit{Kilonova Seekers} on 2023 July 11, over a $\sim$ 6 month period until the end of the O4a observing run of the LIGO-Virgo-KAGRA (LVK) gravitational-wave detectors, on 2024 January 16.  As the primary aim of GOTO is to follow up gravitational-wave alerts from LVK, the timeframes for \textit{Kilonova Seekers} are strongly driven by the schedules of these observing windows. In \sref{GOTO} we begin by introducing GOTO and the need for a citizen science project. In \sref{Workflow} we discuss the \textit{Kilonova Seekers} project in terms of the data used, the workflow and interface the volunteers interact with, the behind-the-scenes machinery, and the alerting and reporting mechanisms. We present in \sref{Volunteers} statistics about volunteer classifications, demographics and engagement, with a particular focus on the valuable contribution of our \lq{power users}\rq. In \sref{ScienceHighlights} we highlight the key scientific results and discoveries from the project, the overall performance of volunteers, and measure the selection function of the volunteers compared to the GOTO real-bogus classifier. Finally in \sref{Conclusions} we summarise the project so far and our key findings, noting our future plans for the project throughout the lifetime of the GOTO survey. A full list of the citizen scientists who were involved with \textit{Kilonova Seekers} can be found in \aref{all_volunteers}.
\section{The Gravitational-wave Optical Transient Observer (GOTO)} \label{GOTO}

GOTO~\citep{Steeghs2022,Dyer_2022} is a multi-site, wide-field telescope array designed to observe electromagnetic counterparts to gravitational wave events – specifically the afterglow of compact binary mergers involving a neutron star, known as kilonovae. GOTO operates in two distinct observing modes: "triggered follow-up" and "all-sky survey" \citep[see][]{Dyer2020}, to rapidly target and tile over the regions associated with incoming alerts, such as gravitational-wave alerts from the LIGO-Virgo-KAGRA (LVK) detectors. While other transients, such as supernovae, take a few weeks on average to reach their optical peak brightness \citep{Anderson2014,Taubenberger2017,Perley2020}, kilonovae peak around 1 day after merger \citep[e.g.][]{Li1998,Kasen2013,Arcavi2017}. Surveys optimised to find kilonovae must have quick responses to alert triggers, fast survey cadence, and efficient transient identification methods. GOTO's overall field of view is larger than the localisation skymap of GW~170817 \citep{Abbott2017}, the only gravitational wave (GW) event with a detected electromagnetic (EM) counterpart, and can cover the whole sky in 2-3 d -- so is ideally suited for these types of searches. 

Due to a combination of the large sky coverage and fast cadence in all-sky survey mode, GOTO collects and generates large volumes of data (500 GB/24h raw, 2-5 TB/24h dataproducts) that make unfiltered human vetting challenging. To address these data volumes, GOTO uses a real-bogus classifier (\texttt{gotorb}) based on a convolutional neural network (CNN) to classify candidate transients in difference imaging \citep[for more information, see][]{Killestein2021}. Each classification is given a probability of being real, and an associated confidence level between 0 and 1. This classifier is effective at filtering out bogus detections, with a 97 per cent recovery rate of real transients for a fixed false positive rate of 1 per cent. As seen with other citizen science projects such as \textit{Supernova Hunters} \citep{Wright_2017}, CNNs and human classifiers have different strengths, which when combined can make a more efficient process than only using one. CNNs are very good at processing large volumes of data, and human classifiers perform better than CNNs when the image is more ambiguous, and when there are not many examples to compare it to. 

\section{The Citizen Science Platform} \label{Workflow}
Given the significant volumes of detections generated, only the highest-scoring candidates from a gravitational-wave follow-up can be prioritised for eyeballing by the GOTO collaboration. By the imperfect nature of classification algorithms, a number of false negatives will always exist below the chosen score threshold, potentially being astrophysically interesting. By lowering the score threshold, we can improve recovery rates, although naturally with increased false positives.

Beyond the real-time necessity for fast transient searches, increasing the possible size of human-labelled datasets is important for training improved classification algorithms. The presence of \textit{label noise} (inaccurate labelling, see e.g. \citealt{frenay2013classification}) is a strong limiting factor in pushing accuracies from $99\%$ to $99.9\%$ (and beyond) and can likely only be mitigated via grouping of labels, weighting by quality of data item, and clipping of bad or unrepresentative examples. 

Citizen science provides a methodology to scale data labelling tasks from small teams of expert scientists, up to thousands of individuals. Calibrated uncertainty quantification is also a crucial missing link in many current astronomical classifiers  (e.g. \citealt{Abdar2020}). Although strides with Bayesian neural networks (e.g.~\citealt{ValentinJospin2020}) have neatly quantified uncertainties associated with choice of model, this often does not represent the uncertainty (or confidence) a human would assign to their prediction. The \textit{true} nature of uncertainties in ML is a complex issue, however, nominal estimates are useful in active learning (where models may suggest which data is most informative to be labelled by a human, e.g. \citealt{Ren2020}), anomaly detection, and decision making rules under uncertainty.

Given these challenges, a citizen science approach is well-suited to generating the scale (and quality) of labelled datasets required to train improved classifiers, and drive searches for candidates that may otherwise be missed in real-time. \textit{Kilonova Seekers} launched in July 2023, after a short beta-testing period with live volunteers. At its core, \textit{Kilonova Seekers} streams uncurated difference image detections (referred to as `candidates' herein) meeting certain cuts from the GOTO real-time pipeline (see Lyman et al., in prep.) to the Zooniverse platform, populating a workflow with pre-baked data visualisations (known as subjects) to receive annotations and classification from citizen scientist volunteers. Through custom infrastructure (see \sref{sec:workflow}), we listen to the classification stream from Zooniverse in real-time, and use this to trigger alerts according to set rules on consensus. We elaborate further on the specifics of this process in the following sections.

\subsection{Data extraction, pre-processing, and presentation} \label{data}
\textit{Kilonova Seekers} ingests candidates as part of a scheduled task -- executed on a daily cadence during project launch, and increased to every three hours during the O4a observing run. Given the multi-site nature of GOTO, this leads to 8 uploads of data per day (weather-permitting). A candidate corresponds to a single difference image detection -- analogous to the concept of \textit{alerts} in other transient surveys. For logistical reasons, \textit{Kilonova Seekers} does not take into account multiple candidates at the same location being associated (i.e. operating at a source level) -- which would require more complex logic to de-duplicate candidates, adding additional overhead. This is intentionally decoupled from how candidates are handled internally, to provide an independent dataflow.

The numbers of real transients and artefacts are heavily imbalanced \citep{Bloom2012}, thus we sample difference image detections uniformly in their real-bogus score (with values between 0 and 1 inclusive, see \citealt{Killestein2021}) through a process of histogram equalisation -- selecting a uniform number of candidates per real-bogus bin, with typical equal bin-size of 0.1. Although these choices necessarily bias the dataset generated, there still exists sufficient diversity to re-balance (and thus train classifiers on) the final dataset.

Candidates are queried from the main difference photometry table generated by GOTO's \texttt{kadmilos} data processing pipeline (see Lyman et al., in prep.), up to a user-specified maximum to avoid flooding volunteers with candidates in the case of rich fields. A number of operational considerations drive the exact query used to ingest candidates -- with our selection cuts being:
\begin{itemize}
    \item Signal-to-noise greater than 10: to minimise the number of false alarm detections due to correlated noise in the initial stages.
    \item Avoidance of the Galactic plane ($|b| < 10^{\circ}$): to minimise the number of variable sources being uploaded to \textit{Kilonova Seekers} -- both for practical rate-limiting purposes, as well as dataset imbalance considerations.
    \item Exclusion of specific GOTO unit telescopes (UTs): owing to ongoing hardware issues, one specific UT was disabled in the \textit{Kilonova Seekers} live workflow to minimise impact on volunteers.
    \item Cuts on images with extremely high numbers of difference image detections: after excluding the plane, these are likely to be poor subtractions which affect class balance. We impose that number of detections in each difference image must be less than the \nth{90} percentile number of detections across all difference images.
    \item Real-bogus score: for the purposes of fast discovery, we adopt a real-bogus score of 0.7 or greater. This is slightly below the normal score threshold of 0.8 used internally, and corresponds approximately to the equality point betwen false positive rate and false negative rate, a common choice in ML contexts.
\end{itemize}

\begin{figure}
    \centering
    \includegraphics[width=\linewidth]{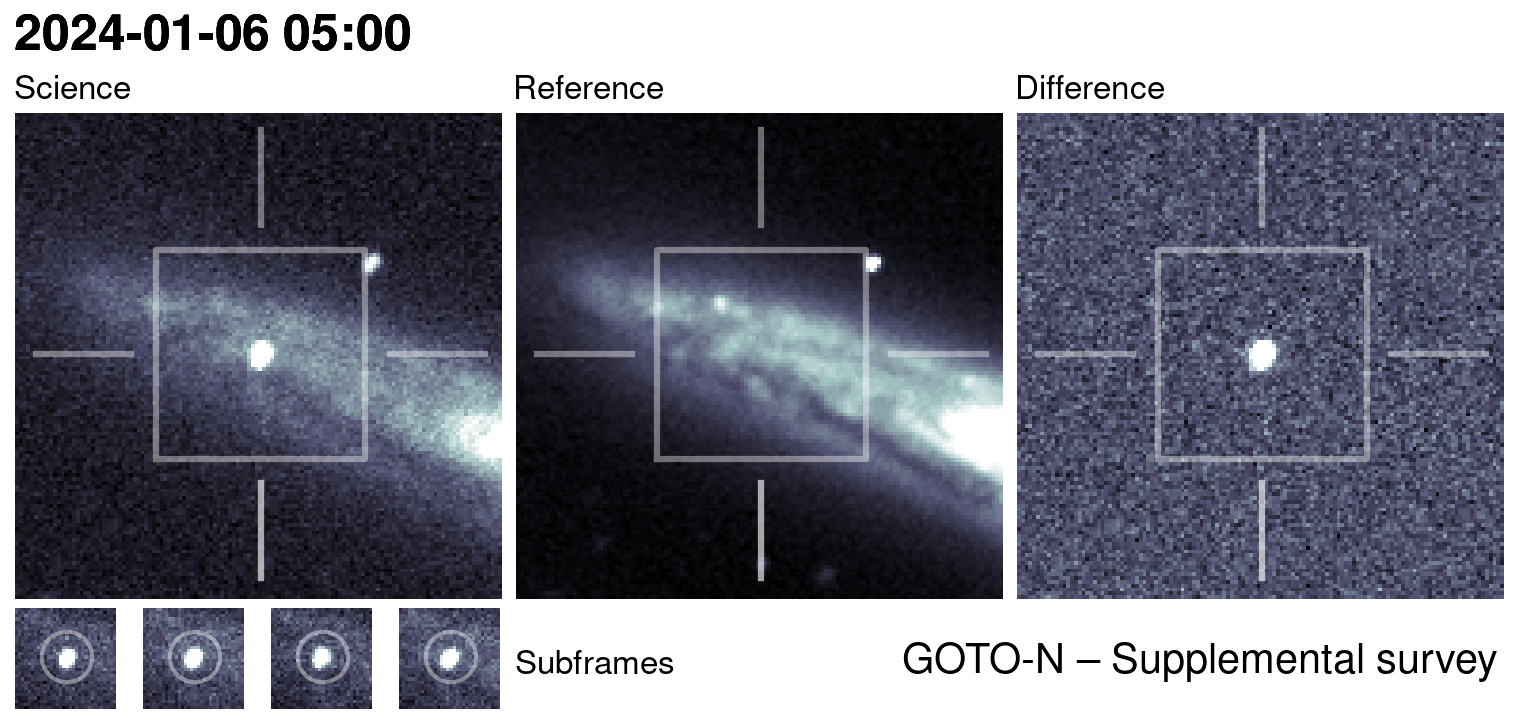}
    \includegraphics[width=\linewidth]{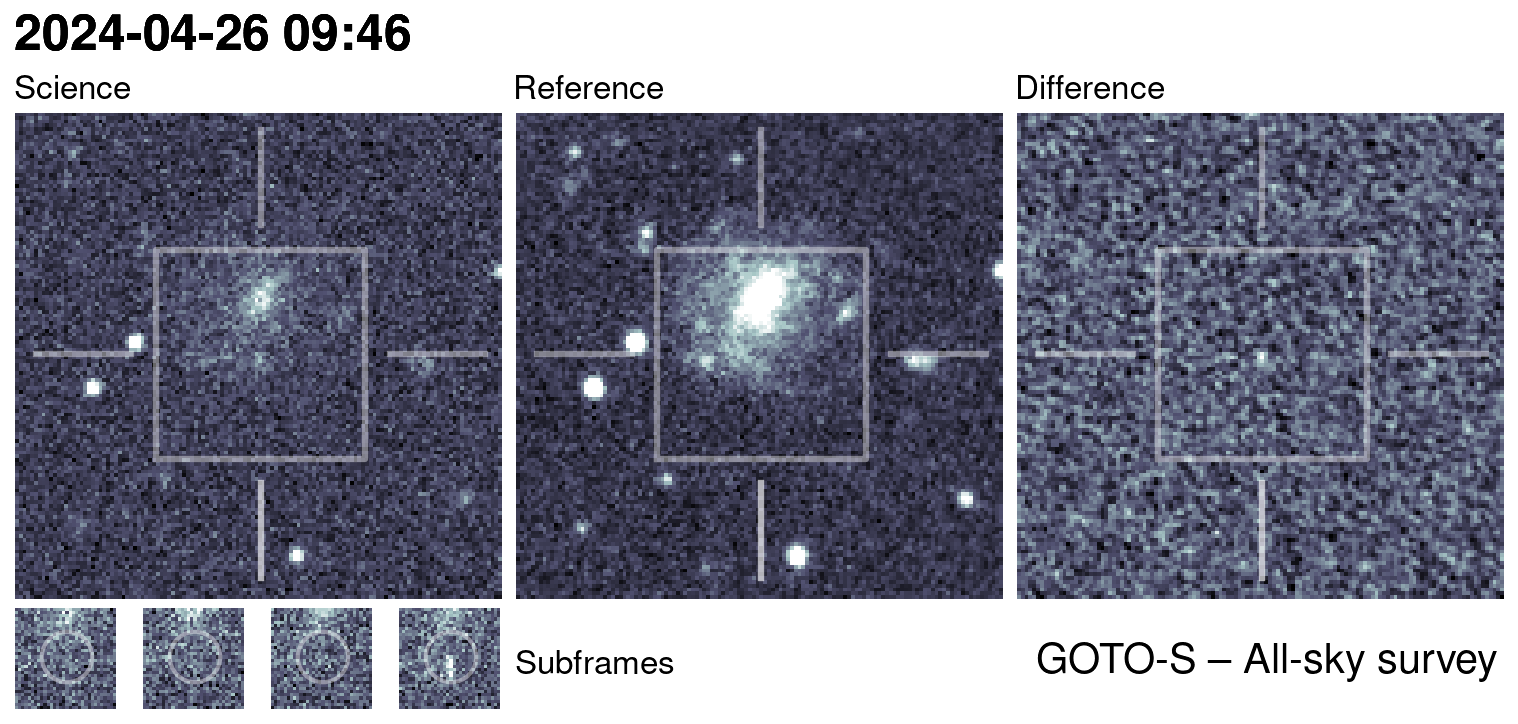}
    \caption{Example subjects from \textit{Kilonova Seekers}. The science, reference, and difference images are plotted, along with subframes and event information.
    The top layout shows SN2024gy, a Type Ia supernova in the nearby galaxy (13.5 Mpc) NGC 4216 flagged by volunteers. The bottom layout shows a cosmic ray artifact that was flagged by volunteers, visible in only one of the four sub-frames, and unfortuitously projected on top of a galaxy.
}
    \label{fig:zooniverse-subject}
\end{figure}

We extract a set of stamps, sized approximately $3\times3$ arcminutes, from the science, reference, and difference images, small cutouts of the main images centred on each candidate detection. The science and reference images are derived from stacked data products, a sigma-clipped combination of a number of individual sub-frames, to reject single-image outliers such as cosmic rays. Stamps are extracted at native GOTO pixel scale (1.4 arcsec/pixel). Pixel thresholds are set using the IRAF \texttt{zscale} algorithm~\citep{Tody1986,Tody1993}, per-channel to span their full range. In a break from the norm of other transient discovery projects on Zooniverse, we use colourised images: specifically the \texttt{matplotlib} \texttt{bone} colourmap. The tasteful blue shading is intended to minimise visual stress. To generate and upload a subject to Zooniverse, we construct a pre-baked layout that we populate with stamps and metadata for a given candidate. We prominently display the detection time into each stamp, to reinforce the real-time nature of uploads to the volunteers, and write which survey each image comes from: to alert volunteers to any images from gravitational-wave (GW), gamma-ray burst (GRB), or neutrino follow-up. The overplotted cross-hairs draw attention to the centre of the frame, and the box shows the field-of-view that the GOTO real-bogus classifier sees, providing important context. We illustrate a subject in \fref{fig:zooniverse-subject}.

Early in Gen. 1 \textit{Kilonova Seekers}, we noticed volunteers overwhelmingly classifying cosmic rays (CRs) as real detections, in spite of their often non-PSF-like appearance and documentation on the field guide for these objects. This motivated the addition of the \textit{subframes} panel for Gen. 2 -- in which we display the individual images that compose the stack -- to identify single-frame artifacts such as CRs that propagate into the stack. These are visible in \fref{fig:zooniverse-subject}, and are $32\times32$ pixels each, with a faint circle added to aid the user in identifying potential moving targets. 

Based on feedback from the volunteers, we added labels to show the volunteers which GOTO site the data originates from, and an event tag to explain which mode GOTO was in when the image was taken. As GOTO is focused on transient follow-up, driven by triggers from external facilities -- the types of images that the volunteers are presented with may change on a daily basis. For example, in survey mode many galaxies may be present in the images, whereas if GOTO is following a specific alert, the telescopes may be pointed towards regions of greater source density, with images being dominated by nearby variable stars in our galaxy. To explain this clearly to our volunteers, we use the following event labels and provide links to the individual instruments listed here so that they can find more information if they are interested in learning more:
\begin{itemize}
\item \texttt{All-sky survey} - GOTO is scanning the sky systematically to find new sources.
\item \texttt{LVK alert [alert number]} - GOTO is following a specific gravitational-wave alert from the LIGO-Virgo-KAGRA (LVK) detectors, searching for the potential optical counterpart.\footnote{\url{https://emfollow.docs.ligo.org/userguide/}}
\item \texttt{Fermi alert} - GOTO is following a GRB alert from the Fermi Space Telescope.\footnote{\url{https://fermi.gsfc.nasa.gov/}}
\item \texttt{Swift alert} - GOTO is following a GRB alert from the Swift Space Telescope.\footnote{\url{https://swift.gsfc.nasa.gov/}}
\item \texttt{IceCube alert} - GOTO is following a neutrino alert from the IceCube detector.\footnote{\url{https://icecube.wisc.edu/science/icecube/}}
\item \texttt{Supplemental survey} - GOTO is doing something else that isn’t covered by the other event tags.
\end{itemize}

Some metadata is deliberately censored from the volunteers, such as the sky location of each candidate, and exact discovery time. This is predominantly to prevent volunteers from seeking additional contextual information outside of the image, that would e.g. confirm a given detection is a minor planet and thus real, as well as for operational reasons to prevent any discoveries being correlated with GW event skymaps, or reported without scrutiny on TNS or social media channels. This policy will naturally evolve with workflow requirements, with in-development workflows (see \sref{Conclusions}) providing additional (albeit carefully chosen) contextual information for classifications.

\subsection{Workflow and ingestion} \label{sec:workflow}
\textit{Kilonova Seekers} presents one unified workflow to the user, tailored to the real-bogus paradigm for source classification. Subjects are shown to volunteers randomly, from the pool of data that has not reached retirement (when voted upon by 15 volunteers). Volunteers are asked if a real source exists at the centre of the crosshairs in the science and difference images. Initial beta tests including a fuzzy \textit{maybe} option showed volunteers overwhelmingly ($\gtrsim 50\%$) selected this option, hindering consensus estimates and making uncertainty estimation impossible.

\begin{figure*}
    \centering
    \includegraphics[width=\linewidth]{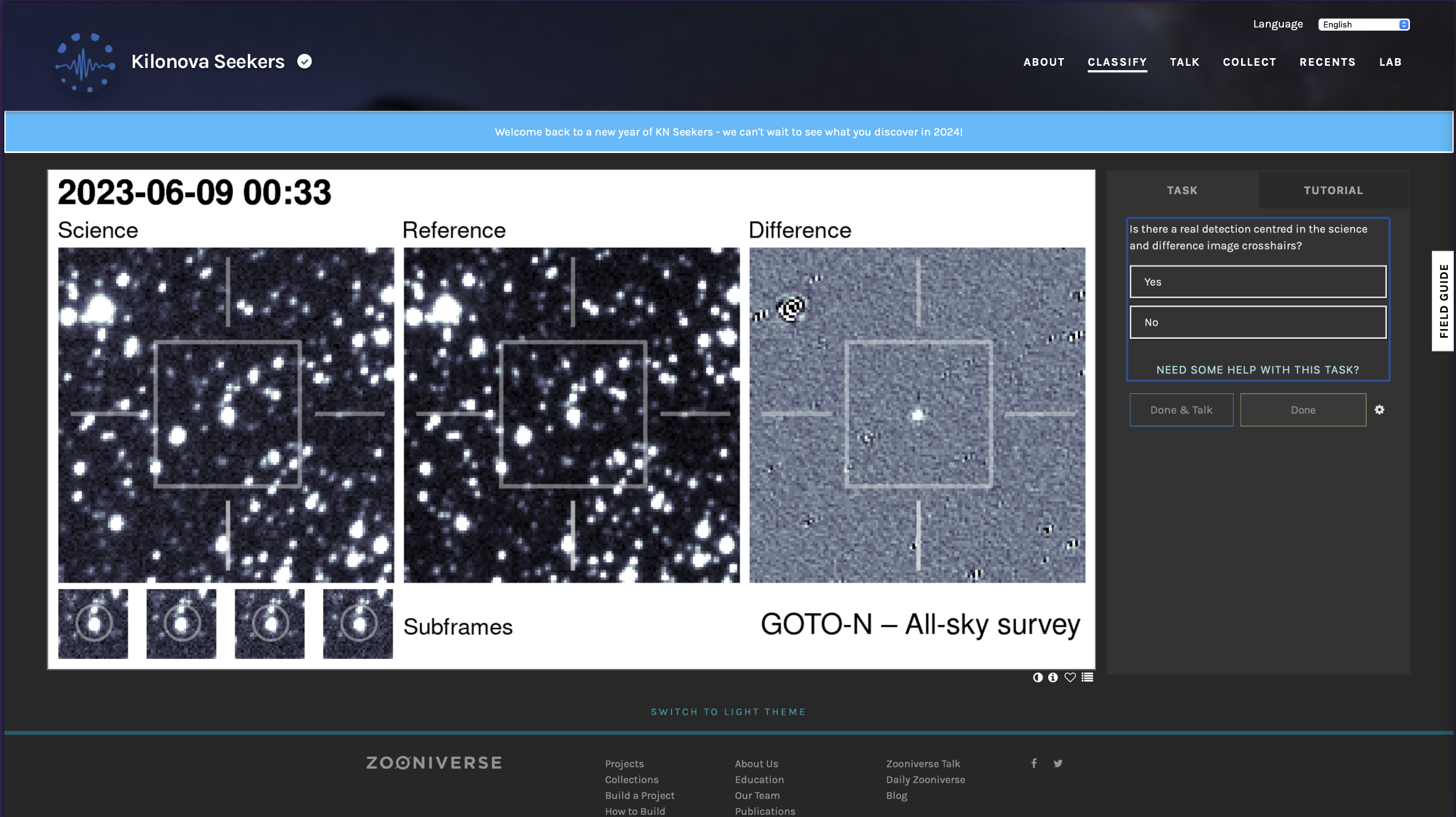}
    \caption{Screenshot of the live \textit{Kilonova Seekers} main workflow.}
    \label{fig:workflow_mainscreen}
\end{figure*}

The web workflow is depicted in \fref{fig:workflow_mainscreen}.
\textit{Kilonova Seekers} also has a companion mobile workflow, delivered via the Zooniverse app. This has the same layout as the web workflow, but with the addition of an intuitive `swipe left and right' interface familiar from other popular mobile apps. We defer a full discussion of the workflows and their utilisation to \sref{demographics}.

\subsection{Alerting and reporting} \label{alerting}
Alerts are intended to flag an object for further follow-up once a given candidate (subject) reaches a configurable consensus threshold. For \textit{Kilonova Seekers} this is set at a threshold of 80\% agreement, and a minimum of 8 votes for the majority option -- set through empirical testing during beta. The high minimum vote threshold is crucial to avoid false consensus, where the wrong answer may be locked in by an early run of votes. This was determined empirically, but is further motivated statistically by ensuring an error of $\sim$10\% in the derived agreement fraction.

\begin{figure}
    \centering
    \includegraphics[width=\linewidth]{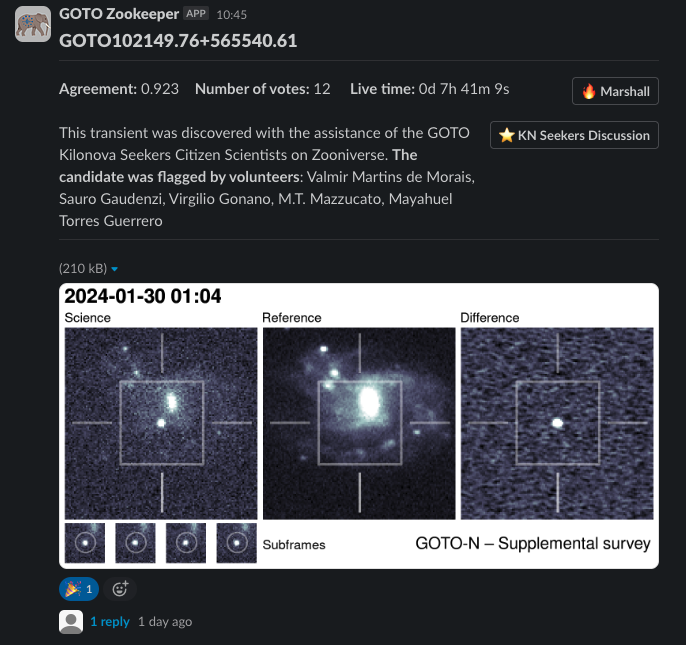}
    \caption{Alert card for a \textit{Kilonova Seekers} candidate that has reached consensus, published via Slack. Visible on the alert card are the consensus level for the candidate, links to both internal GOTO webpages and the \textit{Kilonova Seekers} discussion forum, and the candidate itself.}
    \label{fig:knseekers-alert-card}
\end{figure}

Alerting to the collaboration is delivered via Slack\footnote{\url{https://slack.com}} (the communication platform used by the GOTO collaboration), using the Incoming Webhook API to post an alert card to a dedicated \texttt{\#knseekers-alerts} channel for rapid triaging of candidates.
One such alert card is displayed in \fref{fig:knseekers-alert-card} -- with action links to direct the vetter to the internal GOTO Marshall (see Lyman et al,. in prep.), a web interface for further analysis of transients and reporting, or to the \textit{Kilonova Seekers} Talk pages to check discussion on the object. Collecting key information via a collaborative platform provides a way to centralise discussion about candidates in a maintainable, open way.
Real extragalactic transients are reported to the Transient Name Server (TNS\footnote{\url{https://wis-tns.org}}) through the existing GOTO Marshall architecture. To credit volunteers for their work, we append the names of 5 randomly-selected classifiers of a given transient to the TNS remarks section, subject to integrity checks (see \sref{Volunteers}). This randomisation occurs at point of consensus, and is done in this way to more fairly assign credit, rather than just the first (who may be in a more favourable time-zone, for example). Regardless of this prompt report, all volunteers who correctly identify a given transient are credited on the project results page.

\subsection{Implementation details}
To power the real-time nature of \textit{Kilonova Seekers}, we developed a web service to receive classifications from Zooniverse in low-latency (typically in $\sim$s), combine them with contextual information from the GOTO Marshall, and generate alerts for promising transients.

We use Zooniverse's Caesar\footnote{\url{https://github.com/zooniverse/caesar}} tool to generate a stream of classifications, pushed into a PostgreSQL database hosted locally via a HTTP POST endpoint, exposed on the database machine. The web endpoints for \textit{Kilonova Seekers} are write-only by design, delivered via Apache2 backed by the Python \texttt{django} framework. Schema validation via \texttt{pydantic} ensures only POST requests containing valid classifications are ingested, and enforces strong type safety by checking and enforcing that ingested data are of the right type, enhancing reliability. As Zooniverse predominantly use NoSQL databases internally and make heavy use of free-form JSON data throughout their APIs, we make no attempt to normalise these at ingest and instead use PostgreSQL's excellent native support for JSON(B) datatypes, despite it being a relational database at heart. This was largely driven by the requirement for the database to host ingests from multiple projects, including the internal GOTOzoo project used for GOTO template vetting. Given that different projects may have different metadata (provided as JSON strings), we create project-specific database views for each project, to ensure queries can be written in simpler, more user-friendly ways, without having to parse the JSON strings each time.
The full \textit{Kilonova Seekers} database and real-time stack is hosted on low-power commodity hardware, specifically a cloud-hosted Raspberry Pi Model 4B. Although comparatively tiny, we found this hardware performed ably throughout the first 6 months of the project with over a 99.9\% uptime -- proving highly capable and handling peak throughputs of $\sim$100 classifications per second during the initial launch rush phase. We are currently in the process of migrating \textit{Kilonova Seekers} to more powerful hardware, as we introduce active learning and online ML estimators to our workflows, though this is predominantly for operational stability and could easily remain in-situ. To provide monitoring of the health of the project, Grafana\footnote{\url{https://grafana.com/}} and Prometheus\footnote{\url{https://prometheus.io/}} are used to construct real-time dashboards to visualise the rates, ratios of real-bogus, and bulk properties of incoming classifications. Metrics such as the daily number of active users and classification rate are crucial for informing ongoing engagement strategies and thus are prominent in the design.

We anticipate open-sourcing various aspects of the real-time flows of \textit{Kilonova Seekers} in the near future, to enable the community to make use of pre-built utilities for real-time citizen science projects -- especially in light of new transient surveys coming online in the near future that aim to deliver citizen science components, for example the Vera C. Rubin Observatory (e.g.~\citealt{Higgs2023}).
\section{Volunteers} \label{Volunteers}
As a citizen science project, our Zooniverse volunteers are the key to the success of \textit{Kilonova Seekers}. For us, it is not only important that the project provide useful classifications for improving the GOTO real-bogus classifier, but that the volunteers contribute to meaningful scientific discovery, engage with our collaboration and the other volunteers, learn from the project, and most crucially, enjoy participating in the science of GOTO. 

In this section we discuss the volunteer classifications, highlighting the valuable contribution of our most prolific users (in the top 25, herein power users); before exploring the volunteer demographics, engagement, and the speed and efficiency of their classifications. 

\subsection{Volunteer Classifications} \label{volunteer_classifications}
\begin{figure}
    \includegraphics[width=\linewidth]{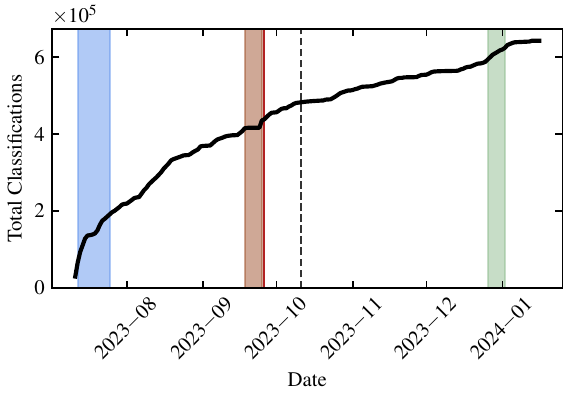}
    \caption{Cumulative classifications per day on \textit{Kilonova Seekers} from launch until the end of O4a (2024 January 16). The blue shaded region corresponds to the dates of press releases, and active media coverage of the project during the launch period. The red shaded region towards the end of September shows the maintenance period after three months of operations, when we temporarily paused the scheduled uploads and implemented the Gen. 2 subjects based on feedback from the volunteers. The green shaded region highlights the increase in rate of classifications over the winter holiday period and the subsequent return to work. The solid red line corresponds to the date of an email newsletter sent out to registered volunteers, leading to a clear increase in classifications. The dashed line is the date we increased the data upload cadence from twice per day to every three hours.}
    \label{fig:classifications_per_day}
\end{figure}

\textit{Kilonova Seekers} launched publicly on Zooniverse on 2023 July 11 at 14:30 UTC, achieving 1000 classifications within the first 30 minutes. 
Coinciding with the project launch, \textit{Kilonova Seekers} was featured in press releases from the GOTO partner institutions and social media, and the \textit{Kilonova Seekers} leads (T.L.K and L.K) were interviewed about the project on the radio for BBC Radio Solent\footnote{\url{https://www.bbc.co.uk/sounds/play/live:bbc_radio_solent}} and on the \lq{Missing Links}\rq\ show on Dublin City FM.\footnote{\url{https://www.dublincityfm.ie/shows/missing-links/}} This period of active publicity is highlighted in blue on \fref{fig:classifications_per_day}, where the impact of this can be seen by a steep gradient in the rate of classifications. 

After the initial launch rush, classifications settled down to an average of $\sim 4000$ classifications per day over the course of the first 3 months of operations.  We consider this time to be \lq{Gen. 1}\rq\ of \textit{Kilonova Seekers}. During this time, only GOTO-North was included, and we were operating the \textit{Kilonova Seekers} project with a once-per-day upload cadence, along with the Gen. 1 image style that did not contain the subframes for easier detection of cosmic rays (as discussed in \sref{data}). As illustrated in \fref{fig:classifications_per_day} by the red shaded region, we paused the scheduled uploads for a week to rapidly implement the Gen. 2 subjects based on feedback from the volunteers, and to upgrade the behind-the-scenes infrastructure ready for ingesting subjects from GOTO-South and the planned increase in upload cadence. We announced our new Gen. 2 subjects in an email newsletter once the maintenance was complete, as indicated in \fref{fig:classifications_per_day} by a solid red line. Classifications quickly increased again to an average of $\sim 3100$ classifications per day after this maintenance period.

GOTO-South at Siding Spring Observatory was integrated successfully into our upload pipeline, and we moved to a three-hour upload cadence on 2023 October 11, as indicated by the dashed line in \fref{fig:classifications_per_day}. Classification rates did slow after this period to an average of $\sim 1700$ per day, however this was largely due to poor weather at both sites due to the changing seasons, meaning there were fewer data to upload to the project.

A particularly interesting feature of \fref{fig:classifications_per_day} is highlighted by the green shaded region. This indicates the Christmas holiday period (December 24 -- Jan 1), when many people are off work for around a week. We found a significant increase in classifications during this time, suggesting that our users may have had more free time to engage with \textit{Kilonova Seekers} - as evidenced by an increase in Talk posts from many of our users during this period. 

In total, over the course of this initial run of \textit{Kilonova Seekers}, between launch and the end of O4a, our volunteers achieved 643,124 classifications of 42,936 subjects.

\begin{figure}
    \includegraphics[width=\linewidth]{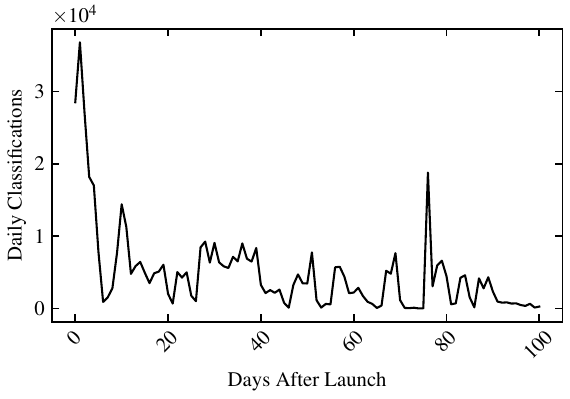}
    \caption{Classifications per day on \textit{Kilonova Seekers} for the first 100 days after launch. This distinct classification curve shows that volunteers regularly classify on the project with the release of new data.}
    \label{fig:dailyclass_100days}
\end{figure}

By focusing in on the first 100 days post launch, we can compare the classification curve of \textit{Kilonova Seekers} (\fref{fig:dailyclass_100days}) with other projects on the Zooniverse. As discussed in \citet{Spiers2019}, the majority of projects on Zooniverse show high classifications on project launch that rapidly declines after the initial launch rush. Occasional peaks in activity may be seen after periods of project promotion, press coverage, or further data release. Other projects such as \textit{Supernova Hunters} show a dramatically different classification curve \citep[see Fig. 4 in][]{Spiers2019}, with more regular spikes in classification indicative of recurring activity. For \textit{Supernova Hunters}, these spikes were on a weekly cadence, resulting from the weekly data upload and newsletter cadence of the project. \textit{Kilonova Seekers} falls somewhere in-between these two trends. The project shows a clear initial launch peak and rapid decline, with smaller regular spikes in activity, likely corresponding to our regular daily upload cadence (barring any weather restrictions).

\subsubsection{Power Users}
\begin{figure}
    \includegraphics[width=\linewidth]{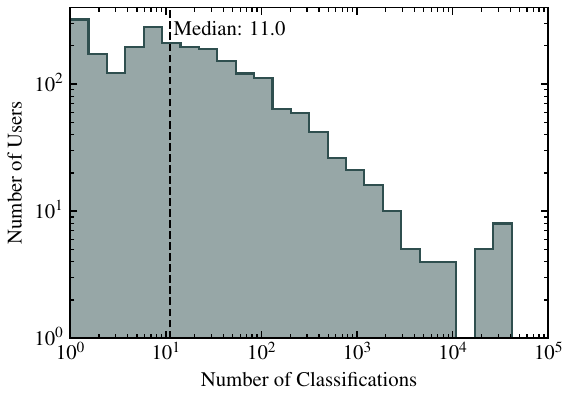}
    \caption{The distribution in classifications among users from launch until the end of O4a. The median number of classifications is 11; however, we have a strong core user-base, with a number of users completing more than 10,000 classifications each.}
    \label{fig:user_classifications}
\end{figure}

\begin{figure}
    \centering
    \includegraphics[width=\linewidth]{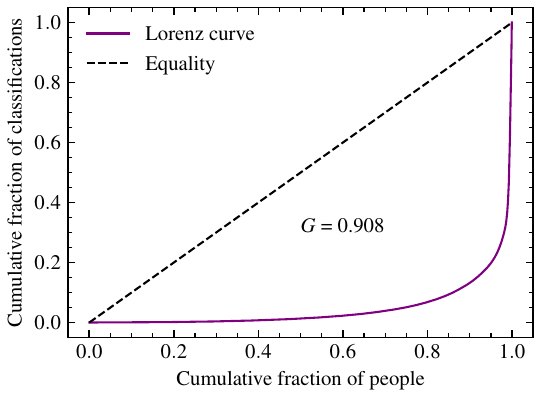}
    \caption{Pareto plot of the cumulative fraction of \textit{Kilonova Seekers} participants from launch until the end of O4a, plotted against cumulative fraction of classifications. The dashed diagonal line represents perfect parity/equality in classification effort per participant. The Gini index is annotated, providing a quantitative measure of the inequality in contribution.}
    \label{fig:gini}
\end{figure}

As shown in \fref{fig:user_classifications}, which shows the distribution in classifications among users, many \textit{Kilonova Seekers} volunteers only undertake a few classifications. Similarly to those for \textit{Galaxy Zoo} \citep{Lintott2008} and \textit{Bursts from Space: MeerKAT} \citep{Andersson2023}, the distribution follows a power law, where the majority of volunteers complete between 1 and 10 classifications on the project, with the number of volunteers declining for larger numbers of classifications. Additionally, this plot clearly shows the significant impact of our \lq{power users}\rq\ who have each contributed thousands of classifications to the project. An alternative framework to look at this is via the Pareto-like \citep[e.g.][]{Lorenz1905,CowellBook} plot in \fref{fig:gini}, where the cumulative fraction of classifiers, and their cumulative share of the classification effort is depicted. Around 90\% of the classifications are performed by 10\% of the volunteers, with a Gini index \citep{Gini1912} of 0.9, in line with other Zooniverse projects of a similar nature (e.g. Table 3 of \citealt{Spiers2019}).

The majority of these power users are the most active participants on the Talk pages, regularly asking questions about the project, sharing their experiences, and providing their thoughts and insights to help others. For the next generation of \textit{Kilonova Seekers} we anticipate appointing and training some of these individuals as moderators to aid in the day-to-day running of the project.

\begin{figure*}
    \begin{subfigure}[t]{.49\linewidth}
        \centering\includegraphics[width=\linewidth]{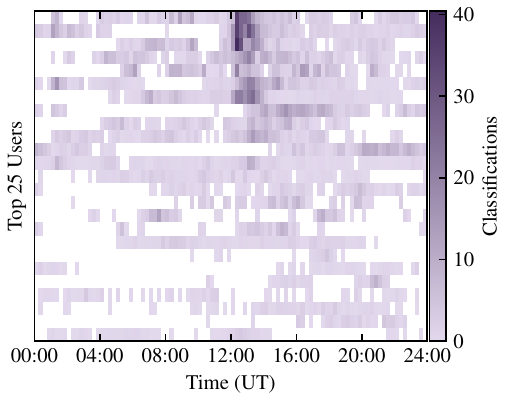} 
        \caption{Average daily classification times for our top 25 users for the time between launch (\nth{11} July 2023) and the \nth{11} September 2023 (a duration of 92 days), separated into 15 minute bins. During this period, new data were uploaded to \textit{Kilonova Seekers} once per day at 12:00 UT.} \label{perday}
      \end{subfigure}
      \hfill
    \begin{subfigure}[t]{.49\linewidth}
        \centering\includegraphics[width=\linewidth]{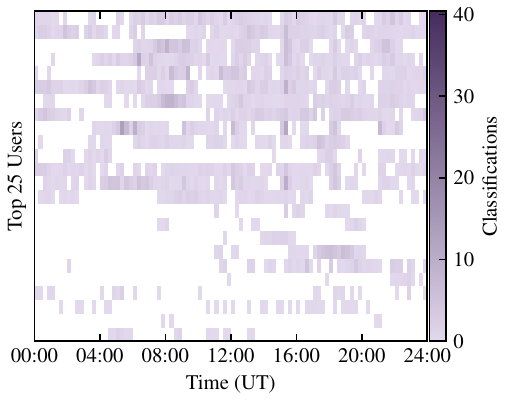}
        \caption{Average daily classification times for our top 25 users for the time between the \nth{11} September 2023 and the end of O4a (\nth{6} January 2024; a duration of 97 days), separated into 15 minute bins. During this period, new data were uploaded to \textit{Kilonova Seekers} every 3 hours.} \label{3hrs}
      \end{subfigure}
    \caption{Average classifications over the course of a day for our top 25 users (as defined by the 25 users with the highest number of classifications between launch and the end of O4a), divided into 15 minute windows. Each row corresponds to a unique user, in descending order to the total classifications over the initial phase of this project, i.e., the top row is the volunteer with the most classifications.}
    \label{fig:users_timeofday}
\end{figure*}

To better understand the classification patterns of the volunteers, we present in \fref{fig:users_timeofday} the average daily classifications for the power user group (the 25 users with the greatest number of classifications between launch and the end of O4a), displayed in 15 minute windows to see trends in volunteer classifications throughout an average day, calculated by dividing the total number of classifications per user per window by the window length in days. We split this into two based on initial daily upload schedule in \fref{perday} and based on the later change to upload new data every three hours in \fref{3hrs}. For the 92 days when we were uploading data every day at 12:00 UT, our most active users were predominantly doing their classifications immediately after the daily data upload. Whilst it is encouraging that volunteers were keen to classify the data immediately, and to be included on the discovery reports, these reports were quickly becoming dominated by the same few volunteers, and others were missing out. This gave further motivation to move to a more frequent data upload -- alongside a more real-time data stream being beneficial for classification speed and distributing the work more fairly. Uploading data more frequently enables volunteers across different timezones to see the data first: allowing them to participate in discovery, and be acknowledged on discovery reports. As illustrated in \fref{3hrs}, during the period where the data were uploaded every three hours, whilst the times that specific volunteers made no classifications remained consistent, there were no longer clear times when the most prolific volunteers did the majority of their classifications.  In spite of these changes, some volunteers still seem to consistently work non-stop on the project, with gaps in \fref{3hrs} likely arising from binning/finite sampling.

\subsection{Volunteer Demographics} \label{demographics}

\begin{figure*}
    \includegraphics[width=\textwidth]{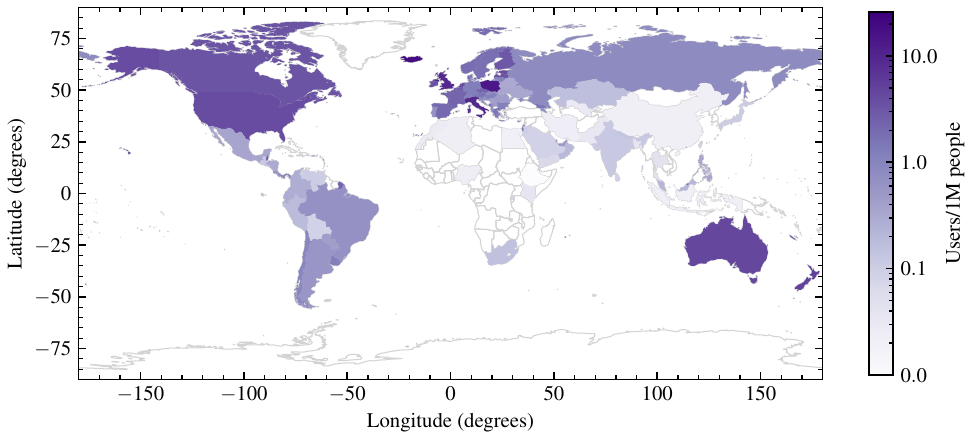}
    \caption{Geographical distribution of volunteers on the \textit{Kilonova Seekers} project. The intensity of a given country corresponds to the classifiers per capita, using information from Natural Earth\protect\footnote{\url{https://www.naturalearthdata.com/about/}}, log-normalised for visualisation purposes.}
    \label{fig:volunteer_map}
\end{figure*}

To date, \textit{Kilonova Seekers} has attracted roughly 2000 volunteers, in over 20 distinct time zones, across 105 different countries. \fref{fig:volunteer_map} displays the geographical distribution of volunteers on \textit{Kilonova Seekers}, shaded according to classifiers per capita. Based on data obtained from Google Analytics, we have participants from every continent (except Antarctica). The wide accessibility of Zooniverse projects enables us to reach countries that may be traditionally underrepresented in astronomical communities. 

Based on the number of users per country, the United States is by far the largest contributor to \textit{Kilonova Seekers}, with a total of 1284 users. At approximately half this value with a total of 615 users is the United Kingdom. However, considering average page views per user for individual countries in the time between launch and the end of O4a, we find that Portugal contains the most prolific \textit{Kilonova Seekers}, with over 2750 views per user on average. 

\begin{figure}
    \includegraphics[width=\linewidth]{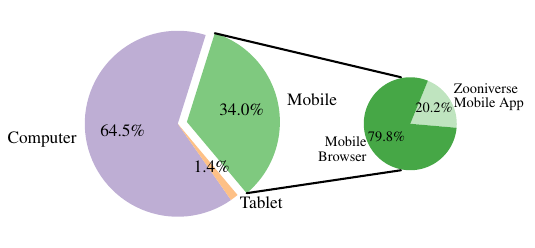}
    \caption{Pie charts illustrating the different ways classifications are made on \textit{Kilonova Seekers}. The larger pie chart indicates the percentages of classifications during O4a that were completed on computers, mobiles and tablets. The smaller, nested pie chart indicates the percentage of mobile classifications done via a mobile browser or the Zooniverse mobile app.}
    \label{fig:userPieChart}
\end{figure}

\begin{figure}
    \includegraphics[width=\linewidth]{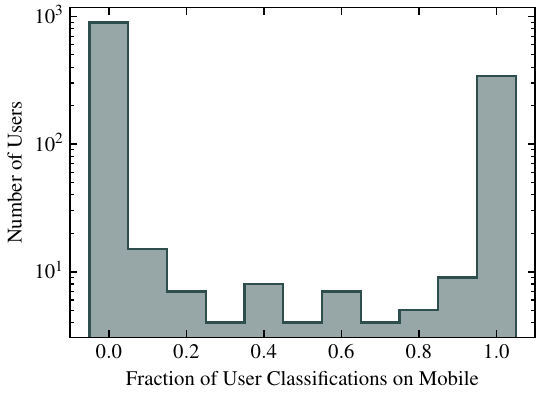}
    \caption{Distribution of the fraction of the total classifications per user performed on a mobile phone. This takes into account both mobile browser and mobile app classifications.}
\label{fig:fracMobile}
\end{figure}

\textit{Kilonova Seekers} is available to all users who can access the Zooniverse platform on the internet, which is available to computer, tablet and mobile users. Alongside the classic in-browser mode, \textit{Kilonova Seekers} is available via the Zooniverse mobile app, available on both iOS and Android devices. The majority of classifications are done via a computer, indicated by \fref{fig:userPieChart}, but roughly a third of classifications are done via mobile phones (inferred via user agent strings). As displayed in \fref{fig:fracMobile}, the fraction of mobile classifications per user is bimodal, with the vast majority of volunteers either not using a mobile phone at all or solely using their mobile phone to engage with \textit{Kilonova Seekers}. Owing to this clear split in our user-base, it is important that future iterations of \textit{Kilonova Seekers} (and other Zooniverse projects) do not contain too many images per page, to ensure continued readability on smaller mobile phone screens.
Although the number of classifications specifically done via the mobile app is relatively small compared to those who use an internet browser (as indicated by the smaller pie chart in \fref{fig:userPieChart}), it represents a non-negligible proportion of participants, necessitating that \textit{Kilonova Seekers} remains compatible with the app, regardless of future updates, so that it remains accessible to all users.

As GOTO is a global collaboration with members from all across the world, it was important to offer \textit{Kilonova Seekers} in the variety of languages that are spoken by the collaboration. At time of writing, \textit{Kilonova Seekers} is available in English, Dutch, Spanish, and Indonesian. We were the first project on the Zooniverse platform to offer Indonesian, and are currently working on the Finnish, Japanese, Polish and Swedish translations, to be released in the near future. However, discussions on the Talk boards predominantly occur in English. These localisations are a volunteer effort driven by GOTO collaboration members, and thus we aim to scale up to support more languages as capacity/enthusiasm allows. 

\subsection{Volunteer Engagement}
The \textit{Kilonova Seekers} team and the wider GOTO collaboration interact with the volunteers via the project \lq{Talk}\rq\ boards, a series of forum pages separated into categories and threads for different discussions. We encourage the volunteers to discuss subjects that they may be unsure of on their individual talk pages, and to ask the GOTO scientists questions by creating their own discussion threads. We use this platform as a key page for announcements to the volunteers from the \textit{Kilonova Seekers} team, including details about new discoveries that they have made and updates about the project or status of the GOTO telescopes. Volunteers can "@" members of the \textit{Kilonova Seekers} team on the Talk pages in the same way as popular social media platforms to alert them if they have a question or need help, and can also send private messages to the team and other volunteers. Through this, volunteers have told us how they have shared \textit{Kilonova Seekers} with their families, friends, amateur astronomy groups, and have discussed the project in blogs and at conferences, widening the overall participation of the project.

On the project Talk pages, volunteers are able to tag their comments. Without any prompting from the team, volunteers started using very similar or the same hashtags as each other. Most of these indicate potential transients with tags such as \texttt{\#real} or \texttt{\#transient}, or highlight other astronomically interesting objects that are not part of the aims of the project e.g., \texttt{\#comet}. The volunteers also use these tags to indicate common artefacts from the field guide, e.g., \texttt{\#badsubtraction} and \texttt{\#satellite}, along with artefacts they have encountered from prior similar citizen science projects, amateur astronomy, and even new ones of their own naming, which we have been able to use not only in our regular field guide updates, but also to update the GOTO hardware team on potential issues. For the next generation of \textit{Kilonova Seekers}, we plan to implement a new multi-class workflow, and these tags will form the basis for the different labels we will include.

Alongside the Talk pages, we engage our volunteers using newsletters. These provide an opportunity to update the volunteers on the status of the project, announce key findings, inform volunteers of changes to the project, and generally share our enthusiasm with the citizen scientists. We have found these to be particularly useful for re-engaging volunteers who may have lost interest in the project over time, as can be seen in the upturn in classifications after a newsletter in \fref{fig:classifications_per_day}.

To ensure that volunteers are credited appropriately for their contributions, discoveries are reported via a dedicated \textit{Kilonova Seekers} results page, including the names or usernames of all of the volunteers who marked a candidate as \lq{real}\rq. Furthermore, we randomly select a subset of 5 names from the \lq{real}\rq\ list to add in a dedicated acknowledgement in the remarks section of the Transient Name Server (TNS) page for the object. In order to receive credit, volunteers must be logged into their Zooniverse account when they make the discovery, so that they can be identified. When volunteers sign up to the Zooniverse platform, they have the option to give their real name. If they have chosen to provide this, their real name will be used for credits, otherwise we use their public username. We automatically filter out email addresses and web links from these text strings.
\section{Scientific Highlights} \label{ScienceHighlights}

In the six months between launch and the end of O4a, the \textit{Kilonova Seekers} project reported a total of 29 objects to the Transient Name Server, which are listed in \tref{table:discoveries}, where 20 of these were official discoveries, first made by \textit{Kilonova Seekers}. 

At present, the candidates that are flagged as interesting by the volunteers require cross-checking by the GOTO collaboration via the Slack alert cards (see \sref{alerting}). Real discoveries are then reported through the TNS via the GOTO Marshall. Anything that is a new discovery and has not appeared yet on the TNS with another group is immediately reported, but \textit{Kilonova Seekers} candidates first identified by other groups are not yet routinely reported owing to limited person-power -- something planned to improve via automation in future updates.

To date, 6 of the 20 transients first discovered by \textit{Kilonova Seekers} during O4a have been classified spectroscopically. The first, AT~2023rob, was classified as a cataclysmic variable star (CV) by the Spectroscopic Classification of Astronomical Transients \citep[SCAT;][]{2022_SCAT} survey \citep{2023_AT2023rob_CR}. The remaining were all classified as Type Ia supernovae \citep{2023_SN2023yer_CR, 2023_SN2023yrs_ysp_CR, 2023_SN2023aajf_CR,2024_SN2023acla_CR} by SCAT, the extended Public ESO Spectroscopic Survey of Transient Objects \citep[ePESSTO+;][]{2015_PESSTO}, and the Young Supernova Experiment \citep[YSE;][]{2021_YSE}. 

In total over the period discussed in this paper, 1037 spectroscopically confirmed supernovae were reported to the TNS, of which 354 subjects associated with these known SNe were generated for \textit{Kilonova Seekers}, assuming the subjects are associated with SNe using a narrow 1\arcsec cross-match radius. Of these, 259 reached the consensus threshold of 80\% agreement and 8 or more positive votes. This implies a recovery fraction of 72\% across this sample, broadly in line with more in-depth estimates presented in in \sref{sec:validation}. A large number of these transients are detected at low SNR, driving the lower recovery than perhaps anticipated -- this figure increases rapidly with SNR, moving to 82\% at SNR=20, 95\% at SNR=50, and 100\% at SNR=70. In the following subsections, we discuss in depth some of these early results from the \textit{Kilonova Seekers} project.

\afterpage{
\clearpage
\begin{landscape}
\centering 

\begin{tabular}{llllllll}
\hline
TNS Name & GOTO Name & GOTO Discovery Date (UT) & \textit{Kilonova Seekers} Subject/s & TNS Reporting Group & RA/Dec & Type & Redshift \\
\hline
\multicolumn{2}{l}{\textit{Kilonova Seekers} Discoveries} \\
\hline

AT2023pmm & GOTO23yt & 2023-08-05 04:48:55 & 91259701 & GOTO: \textit{Kilonova Seekers} & 02:44:18.422 +14:23:27.51 & & \\
AT2023pof & GOTO23vt & 2023-08-08 02:55:13 & 91223502, 91282780 & GOTO: \textit{Kilonova Seekers} & 19:48:39.623 +00:40:25.99\\
AT2023rob & GOTO23aja & 2023-09-05 21:56:05 & 91624786 & GOTO: \textit{Kilonova Seekers} & 18:55:04.878 -25:42:41.94 & CV\\

AT2023wbu & GOTO23bbl & 2023-10-28 06:09:44 & 92889863 & GOTO: \textit{Kilonova Seekers} & 10:48:51.594 +17:37:33.02 \\

AT2023xnj & GOTO23bia & 2023-11-11 10:45:19 & 93524342 & GOTO: \textit{Kilonova Seekers} & 00:21:31.492 -32:48:20.18\\
AT2023xqf & GOTO23biq & 2023-11-10 10:40:28 & 93597712 & GOTO: \textit{Kilonova Seekers} & 00:03:55.159 -29:35:38.95\\
AT2023xqg & GOTO23bip & 2023-11-12 17:07:37 & 93615033 & GOTO: \textit{Kilonova Seekers} & 10:39:28.016 -39:31:33.69\\
AT2023xqy & GOTO23bjh & 2023-11-13 11:06:02 & 93671156 & GOTO: \textit{Kilonova Seekers} & 23:41:43.058 -34:12:06.46 \\
AT2023ydt & GOTO23blc & 2023-11-18 12:16:31 & 93953774 & GOTO: \textit{Kilonova Seekers} & 02:19:40.742 -48:15:32.90\\
SN2023yer & GOTO23blj & 2023-11-18 20:45:07 & 93965156 & GOTO: \textit{Kilonova Seekers} & 01:21:16.700 +17:12:55.98 & SN Ia & 0.06\\
AT2023yox & GOTO23bms & 2023-11-28 04:54:34 & 94193252 & GOTO: \textit{Kilonova Seekers} & 11:55:51.573 +44:08:05.40\\
AT2023yqr & GOTO23bno & 2023-12-02 10:25:06 & 94310806 & GOTO: \textit{Kilonova Seekers} & 01:14:48.773 -20:59:41.45\\
AT2023yqs & GOTO23bnn & 2023-11-30 11:15:33 & 94310814 & GOTO: \textit{Kilonova Seekers} & 02:08:23.440 -35:04:23.95\\
SN2023yrs & GOTO23bnt & 2023-12-03 13:47:40 & 94322374 & GOTO: \textit{Kilonova Seekers} & 06:26:52.896 -24:36:53.01 & SN Ia-91-bg-like & 0.02331\\
SN2023ysp & GOTO23bnz & 2023-12-03 13:29:44 & 94332759 & GOTO: \textit{Kilonova Seekers} & 06:19:37.294 -29:49:16.56 & SN Ia & 0.09\\
AT2023aagc & GOTO23bus & 2023-12-15 12:28:26 & 94836562 & GOTO: \textit{Kilonova Seekers} & 05:29:37.658 -35:55:16.98 \\
SN2023aajf & GOTO23bwl & 2023-12-17 12:01:54 & 94862495 & GOTO: \textit{Kilonova Seekers} & 04:22:41.484 -51:29:15.63 & SN Ia & 0.0428\\ 
AT2023abdm & GOTO23bzu & 2023-12-17 11:31:43 & 95035983 & GOTO: \textit{Kilonova Seekers} & 03:41:14.308 -48:51:18.08\\
AT2023abdn & GOTO23bzs & 2023-12-24 11:49:37 & 95035974 & GOTO: \textit{Kilonova Seekers} & 05:48:49.179 -24:15:21.60\\
SN2023acla & GOTO24P & 2023-12-26 04:17:46 & 95128349 & GOTO: \textit{Kilonova Seekers} & 12:05:02.450 +01:10:32.95 & SN Ia & 0.06565\\
     
\hline
Reported \\
\hline
SN2023oxc & GOTO23uh & 2023-08-04 22:12:26 & 91273350 & ATLAS & 16:04:31.469 +36:19:00.59 & SN & 0.0434\\
SN2023ver & GOTO23bbc & 2023-10-26 00:45:31 & 92809761 & Pan-STARRS & 03:51:40.274 -00:30:38.95 & SN Ia-91T-like & 0.03\\
SN2023vqn & GOTO23bcc & 2023-10-27 21:31:28 & 92889737 & ATLAS & 22:52:31.726 +18:14:06.46 & SN Ia & 0.07\\
AT2023xig & GOTO23bhy & 2023-11-10 13:39:34 & 93464849 & ATLAS & 04:30:41.258 -39:17:55.73\\
AT2023acdo & GOTO23caa & 2023-12-24 11:59:58 & 95134996, 95190835 & ZTF & 06:04:40.410 -26:38:41.64\\
SN2024gy & GOTO24J & 2024-01-06 05:00:18 & 95413590 & Koichi Itagaki & 12:15:51.290 +13:06:56.12 & SN Ia & 0.00118\\
SN2024hm & GOTO24Q & 2024-01-06 10:25:10 & 95430426 & ATLAS & 03:24:06.521 -38:43:59.42 & SN Ia & 0.067\\
AT2024kh & GOTO24X & 2024-01-06 05:33:21 & 95601222 & ATLAS & 13:16:52.136 +28:06:32.66 & \\
AT2024agm & GOTO24fq & 2024-01-06 05:14:39 & 95974795 & ATLAS & 12:57:38.772 +40:11:57.38 \\
\hline
\end{tabular}
\captionof{table}{\textit{Kilonova Seekers} discoveries reported to the TNS, which were observed by GOTO between \textit{Kilonova Seekers} launch (2023 July 11) and the end of O4a (2024 January 16). We present the TNS name, internal GOTO name, GOTO discovery date, \textit{Kilonova Seekers} associated subject id(s) on Zooniverse, TNS reporting group, transient location, and if known, the classified type and redshift. Redshifts are taken directly from the TNS classification report, but rounded where appropriate.}
\label{table:discoveries}
\end{landscape}
    \clearpage
}

\subsection{Rapid reporting}
One of the key accomplishments to highlight from \textit{Kilonova Seekers} is the speed of classification and consensus from the volunteers. As we have volunteers from around the world, there is almost always someone online looking at the data in real-time, whether uploaded to \textit{Kilonova Seekers} (e.g. \fref{fig:users_timeofday}), or internally within the collaboration. Between 2023 September 11 and the end of O4a, we changed the data upload cadence to the Zooniverse platform to every three hours, and found that the majority of new subjects uploaded were classified before the next data upload just three hours later. 

\begin{figure*}
    \centering\includegraphics[width=\linewidth]{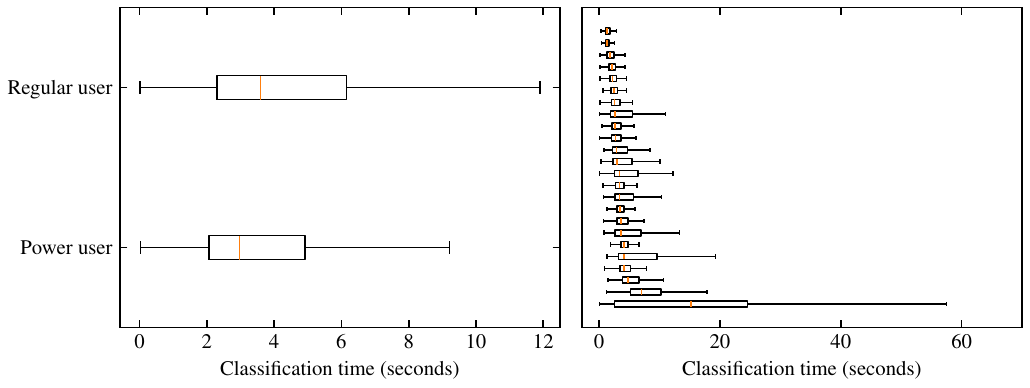}
    \begin{subfigure}[t]{.55\linewidth}
        \caption{Boxplots showing the distributions of classification times of power users, selected as the top 25 most prolific classifiers on \textit{Kilonova Seekers}, compared to the remainder of the user base (regular users).} \label{class}
      \end{subfigure}
      \hfill
    \begin{subfigure}[t]{.44\linewidth}
        \caption{Boxplots showing the distribution of classification times of our 25 power-users, sorted by median classification time.} \label{power_class}
      \end{subfigure}
    \caption{Boxplots showing the classification times of the \textit{Kilonova Seekers} volunteers. Maximum time per classification has been clipped to 2 minutes to remove those classifications where someone paused mid-classification and submitted at a much later time. Orange lines represent the median classification time, the boxes show the upper ($Q3$) and lower ($Q1$) quartile values, with width corresponding to the interquartile range (IQR) and the whiskers represent $Q1 - 1.5\times IQR$ and $Q3+1.5\times IQR$ respectively.}
 \label{class_times}
\end{figure*}

We display in \fref{class_times} the average classification speeds of the \textit{Kilonova Seekers} volunteers per subject. We clip the maximum time per classification to 2 minutes to measure the actual attention paid to the classification -- there were cases where classifications took on the order of 18 hours, which we interpret as situations where a volunteer stepped away from their device and submitted the classification at a later time. As shown in \fref{class}, our power users typically take less time to classify a subject than the remainder of users, who have a wider range of classification times. However, the median classification time for both groups is roughly 5\,s, meaning that if we take our total classifications for the period (see \sref{volunteer_classifications}), our volunteers have dedicated at least 893 hours of classification time to the project during O4a.

In \fref{power_class}, we break down the power-user classification times per user, and explore the distributions. There are clear differences here, with some users routinely taking under 10\,s for every single classification they do, whilst others take substantially longer. This behaviour is unclear, and no conclusive explanation exists. Some power users may be reading and investigating the metadata for the subjects to find more insights that may help them make a classification -- since these attributes are mentioned on the Talk boards by a small subset of volunteers. The final user on the plot is an extreme outlier -- upon detailed inspection this user's classification times show a remarkable bimodality, with a similar `early' peak to the other participants, but with a strong peak around 20\,s, skewing their quartiles on this plot.

\begin{figure}
    \centering
\includegraphics[width=\linewidth]{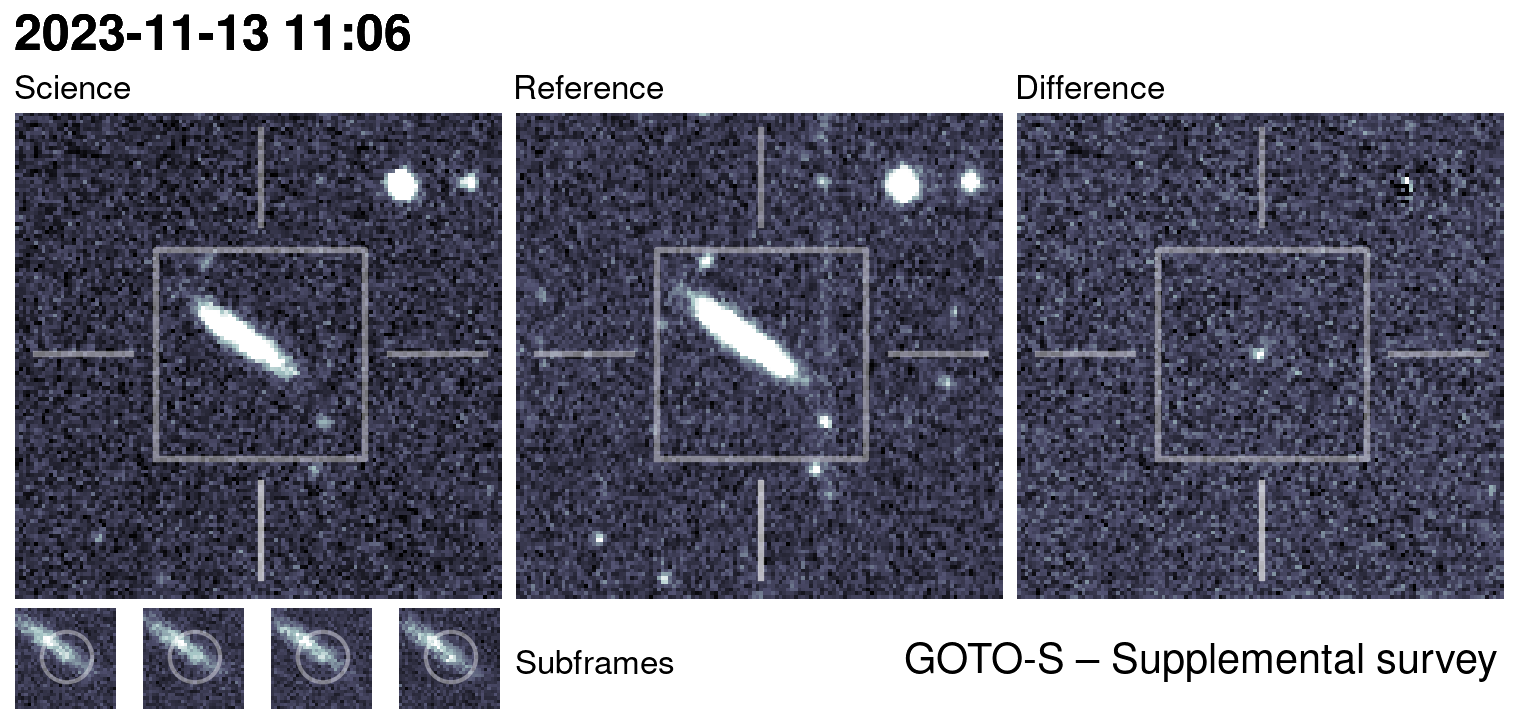}
    \caption{\textit{Kilonova Seekers} subject for AT2023xqy. This transient was flagged by the volunteers as real and reported to the TNS within 3 hours and 20 minutes of data being taken by GOTO.}
    \label{fig:at2023xqy_subject}
\end{figure}

A particularly significant scientific highlight for \textit{Kilonova Seekers} was the discovery of AT~2023xqy (the Zooniverse subject for this discovery is displayed in \fref{fig:at2023xqy_subject}). This object was observed by GOTO-South on 2023 November 13 at 11:06:02.592, and was reported to the TNS at 14:27:36 on the same day. It was observed, the data were reduced and uploaded to Zooniverse, the candidate was flagged as interesting, cross-checked and confirmed as real, and reported to the TNS within approximately 3 hours and 20 minutes of data being taken. This transient had a rapid rise in brightness. The last GOTO non-detection was 24 hours prior at a L-band magnitude of 20.8. The transient was discovered 1 day later at a magnitude of 19.2 -- suggesting this object rose in brightness by 1.6 mag/24\,h, and implying the transient was caught early post-explosion. This finding was later confirmed by ATLAS on 2023 November 17. This speed of human vetting is simply not sustainable without the dedication of our citizen scientists. 

\subsection{Validation dataset, detection efficiency, and volunteer benchmarking}
\label{sec:validation}
Outside of the real-time transient discovery workflow, \textit{Kilonova Seekers} provides a framework for generating a number of human benchmarks, and gold-standard datasets for training machine learning solutions, as a natural byproduct of the transient search workflow. We elaborate on a few ongoing analyses that provide substantial insights into the abilities of our volunteers, and map out the `human factor' present in transient follow-up, that few time-domain projects have previously explored in detail (e.g. \citealt{Goldstein2015,Hayden2021}). To measure the intrinsic performance of volunteers, and determine sensible classification baselines, we inject a number of validation datasets (both intentionally, and intrinsically via known objects) with known answers into the live project:

\begin{itemize}
    \item Hand-labelled validation dataset: 300 examples, sampled uniformly in real-bogus score from detections prior to project launch, and hand-labelled by the Authors to ensure high accuracy.
    \item Minor planets: given the ingest pipeline is agnostic to contextual information, these detections with high real-bogus score naturally enter into \textit{Kilonova Seekers} as part of the transient search workflow. We know \textit{a priori} that these are real detections, and the spatial association enables us to retrieve high confidence low-signal-to-noise detections for testing.
\end{itemize}

The hand-labelled validation dataset is given an arbitrarily high retirement limit to ensure as many volunteers as possible see them for comparative analyses. For the analyses that follow, we neglect the possibility of label noise (inaccurate labelling by the team) in the validation datasets. For the hand-labelled set, these data were vetted by the Authors with both knowledge of the co-ordinates, and additional contextual information (historical variability, source cross-matches) to guide the labelling. For the minor planet dataset, we select only detections with high-confidence ($\leq 4$") matches to catalogued objects from the Minor Planet Centre, following \citet{Killestein2021}.

Through analysis of the validation dataset, and binary classification labels from volunteers, we can assess both the cohort and individual performance of volunteers in a real-world setting. To ensure low sampling noise in our estimations of precision, we only consider volunteers who have completed 100 validation subjects or more, yielding noise of O(1\%). We suspect the validation set size is sufficient to mitigate data-driven scatter in metrics.

\begin{figure}
    \includegraphics[width=\linewidth]{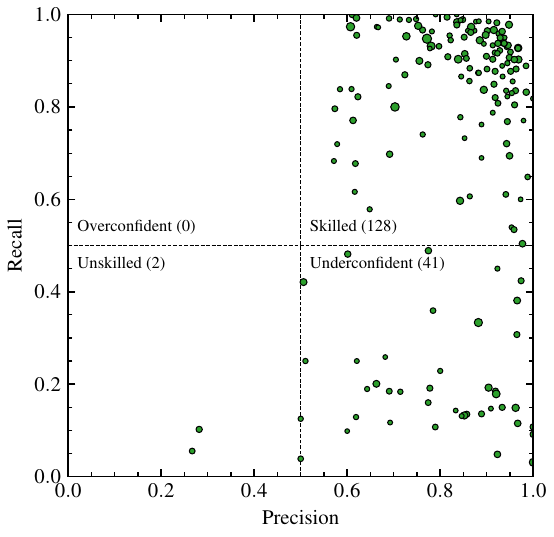}
    \caption{Precision-recall plot for the validation set, computed per volunteer with over 100 classifications. The dashed lines partition the precision-recall space into quadrants, corresponding to the 50\% precision/recall boundary. The size of the plot markers is proportional to the number of classifications performed by that user.}
    \label{fig:recallprecision}
\end{figure}

As shown in \fref{fig:recallprecision}, we plot the precision ($PR$) and recall ($RC$) for each volunteer evaluated on the hand-labelled validation dataset.

\begin{equation}
PR = \frac{TP}{TP + FP}
\end{equation}
\begin{equation}
RC = \frac{TP}{TP + FN}
\end{equation}

where $TP$ is the number of real transients correctly labelled as such by the volunteer, $FP$ is the number of bogus transients incorrectly labelled as real, and $FN$ is the number of real transients labelled as bogus. The $F1$ score is a convenient metric derived as the harmonic mean of these quantities, given as

\begin{equation}
F1 = \frac{2 \cdot PR \cdot RC}{PR + RC}
\end{equation}

where the precision and recall are defined as above.
The volunteers broadly perform well on the validation dataset, achieving a median (class-weighted, 1$\sigma$ uncertainty) F1 score of $78^{+13}_{-35}\%$ and lie in a cluster in the upper right quadrant (precision and recall above 50\%). 
and represents a class-balanced accuracy, weighting precision and recall equally.

There are a notable minority (20\%) of volunteers who lie in the lower right quadrant (high precision, but low recall) -- whom we interpret as `underconfident' volunteers. When they mark objects as real transients, they are likely to be correct, but they mark very few objects as real transients -- perhaps owing to not fully trusting their own predictions. Reassuringly, very few volunteers lie in the low precision region of the plot, characterised by poor discriminative performance -- we associate the upper left quadrant with `overconfident' volunteers, who recover the majority of real transients but mark many artifacts as real. We hope that, over time, volunteers Precision-Recall scores will flow towards the upper right corner as they gain performance and familiarity with the workflow and project.

\begin{figure}
    \includegraphics[width=\linewidth]{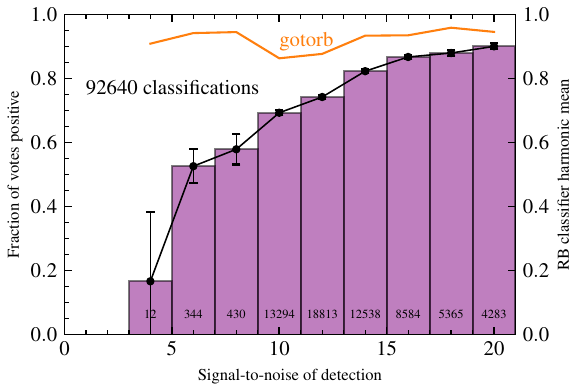}
    \caption{Fraction of positive votes per subject, binned by the SNR of the detection, derived from all live \textit{Kilonova Seekers} minor planet detections. Uncertainties are estimated by the one-sided binomial score interval approximation, with error bars representing 2 sigma. The 50\% recovery threshold sits around signal-to-noise 6. The harmonic mean of the real-bogus classifier score \citep{Killestein2021} per bin is overplotted in orange, for illustration.
    }
    \label{fig:peoplevsrb}
\end{figure}

In \fref{fig:peoplevsrb}, we compare the recovery of minor planets by the volunteers compared to the GOTO real-bogus classifier \citep[see][]{Killestein2021} as a function of the signal-to-noise of the detection. We cross-match all uploaded \textit{Kilonova Seekers} subjects with Minor Planet Centre\footnote{\url{https://www.minorplanetcenter.net}} ephemerides, and in total retrieve 92,640 classifications -- which we know \textit{a priori} are good transient detections. We compute the fraction of positive votes per signal-to-noise bin, chosen approximately to linearly span the range 3 to 20, where the majority of detections typically lie. Uncertainties are estimated from the normal approximation \citep{Wald1943} to the one-sided binomial proportion confidence interval:

\begin{equation}
\sigma_{\hat{p}} = \sqrt{\frac{\hat{p} (1-\hat{p})}{N}}
\end{equation}

which is an adequate and asymptotically-correct estimator, given the typically large $N$ per bin, and lack of bins with $\hat{p}$ close to zero or one.

For comparison, we overplot the harmonic mean of real-bogus classifier scores -- the closest analogy to the fraction of votes positive approach we use for volunteer labels.
This is given as 
\begin{equation}
P = \frac{1}{N}\sum_{i=1}^{N} \frac{1}{p_i}
\end{equation}
where $p_i$ is the $i^{\mathrm{th}}$ classifier score in each bin, and $N$ is the total number of subjects per signal-to-noise bin.
This plot highlights facets of the performance of both human vetters and the real-bogus classifier.
The classifier score remains high across the SNR distribution, as expected. The marked bump at low ($\sim 7$) signal-to-noise in the classifier score is likely a result of the steep power-law slope in the magnitude distribution of minor planets -- with many times more small bodies than larger in the training set (see \citealt{Killestein2021}).
The human classifier scores show a smooth sigmoid curve, passing 50\% recovery around a SNR of 6. Uncertainties (given by the error bar) are largely driven by sample size per bin, rather than human-derived uncertainty. The real-bogus classifier score comfortably exceeds the human true positive rate, markedly so at lower signal-to-noise. It is perhaps not surprising that a classifier explicitly trained on minor planets outperforms a naive ensembling of human predictors -- yet to our knowledge this is among the first validations of deep-learned classifiers outperforming human annotators in time-domain astronomy.  We caution that the human-derived fraction of positive votes may not be well-calibrated probabilistically, taking into account discussions on variable precision and recall of volunteers above -- nevertheless via thresholding and consensus these issues may be mitigated.

Optimal schemes for thresholding or weighting (e.g. \citealt{Marshall2016,O'Brien2024}) are left to future publications, though we note that the \textit{uncertainty} is a crucial component of our science aims, and so fraction of positive votes is diagnostic here. With priors on the true/false positive rates per volunteer from the validation set, Bayesian models of annotation (e.g. \citealt{10.1162/tacl_a_00040}) are a promising avenue for deriving well-calibrated and optimal inferences on how likely an object is to be real from volunteer labelling.

Nevertheless, this result underscores that classifier scores alone are not sufficient to fully capture the uncertainty associated with a classification. Subjects that are genuinely challenging in a statistical sense, such as those at low signal-to-noise, should be treated with nuance to avoid over-interpretation. This underscores the necessity of uncertainty quantification in classification

Although early in the project's lifetime, these validation datasets have enabled a number of interesting scientific (and sociological) insights into the way volunteers approach classification tasks, their intrinsic efficiency at recovering transient objects, and the different dispositions of the volunteers to classification. More advanced validation experiments are currently underway -- including injecting augmented variants of the validation set to track the evolving performance of the volunteers between \textit{Kilonova Seekers} generations. One remaining, potentially insightful task is to re-run our validation workflow with GOTO team members to compare and contrast Figs.\ref{fig:recallprecision} and \ref{fig:peoplevsrb}, and measure the selection function of project scientists \citep[similar to the investigation of][for Martian surface feature detection and classification]{Wardlaw2018} -- which could feed into downstream analyses to derive more informed recovery estimates/drive second-looks on more challenging data.

Based on cuts inferred from the validation dataset, we define our gold-standard dataset as subjects with $>80\%$ agreement, and more than 8 positive/negative votes from volunteers. Based on these cuts, we find a gold-standard dataset of 17,682 detections across O4a. This gold-standard dataset is informing the development of the next real-bogus classifier within GOTO, with a more detailed discussion of nuances associated with crowd-sourced training of transient classification models deferred to a future publication.
\section{Conclusions} \label{Conclusions}
In this paper, we have presented the first stage of \textit{Kilonova Seekers}, a citizen science project designed specifically for real-time transient discovery, complementing the unique capabilities of the GOTO survey for gravitational-wave follow-up.

In the period from July 2023 to January 2024, \textit{Kilonova Seekers}:
\begin{itemize}
    \item Achieved 643,124 classifications of 42,936 subjects.
    \item Attracted roughly 2000 volunteers, in over 20 distinct time zones, across 105 different countries.
    \item Reported 29 objects to the TNS, where 20 of these are discoveries first reported by the project. 6 of these discoveries have been classified spectroscopically by other teams.
    \item Achieved turn-around times of as quick as 3 hours and 20 minutes between observation and TNS report, for candidates flagged as interesting by the volunteers.
    \item Created a gold-standard training set of 17,682 subjects for machine learning, with over 80\% agreement among volunteers. 
    \item Measured the detection efficiency of the volunteers at recovering transient sources, and compared this with the existing GOTO real-bogs classifier. 
\end{itemize}

With this initial phase of \textit{Kilonova Seekers}, we have demonstrated concretely that citizen science can work both in real time and low latency -- driving decision-making and discovery on large data-streams. 

\subsection{Recent updates and future work}
For the O4b observing run which is now underway, \textit{Kilonova Seekers} has continued to grow rapidly and transitioned to an augmented hourly cadence upload, to further reduce the latency between discovery, upload, and consensus. This has led to a number of citizen science discoveries within 2 hours of images being taken. We intend to keep shortening this cadence towards zero-delay (uploads simultaneous with pipeline completion), as survey and platform capacity allow. A new injection of unbiased (spanning the full real-bogus range) candidates, which aggressively sample real-bogus scores across the whole range are proving an excellent seed dataset for novel deep-learned classifiers in development. In the time taken to prepare this publication, \textit{Kilonova Seekers} has now reached 31 discoveries and achieved over 1 million classifications from volunteers. A full discussion of this second phase and ongoing discovery is deferred to future works.

Development of the \textit{Kilonova Seekers} workflows continue, with multi-class, context-augmented workflows planned to be released later in 2024. This will enable volunteers to not only classify if a source is real or bogus, but to subdivide each of these classes into morphological types (e.g. supernova, nuclear transient, variable star). This workflow will further support the training of next-generation machine learning classifiers, and enable uncertainty-aware contextual classification. The introduction of this \textit{Kilonova Seekers} Multi-class will mark Gen.~3 of the project, and be accompanied with a re-launch. This development is, of course, in addition to the original fast discovery workflow, to ensure continuity for volunteers and maintain compatibility with mobile app users.

Based on the keen engagement with \textit{Kilonova Seekers}, a number of parallel companion outreach and public engagement projects are under active development: empowering volunteers to do their own transient follow-up efforts with professional telescopes, learn about time-domain astrophysics through observing objects themselves, and generate meaningful scientific outcomes and publications on the objects they have discovered.

The time-domain community are eagerly following up alerts during the LIGO-Virgo-KAGRA O4b observing run, hoping these GW triggers will facilitate discovery of new electromagnetic counterparts.  With the growth of the \textit{Kilonova Seekers} project, this community is now markedly larger.
\section*{Acknowledgements}
We thank the anonymous referee for their insightful comments which helped improve the quality of this manuscript.
TLK acknowledges support via an Research Council of Finland grant (340613; P.I. R. Kotak), and from the UK Science and Technology Facilities Council (STFC, grant number ST/T506503/1).
LK and LN thank the UKRI Future Leaders Fellowship for support through the grant MR/T01881X/1.
EW thanks STFC for support through the grant ST/Y509486/1.
JDL acknowledges support from a UK Research and Innovation Fellowship (MR/T020784/1).
DMS acknowledges support by the Spanish Ministry of Science via the Plan de Generacion de conocimiento PID2020-120323GB-I00 and PID2021-124879NB-I00.
SM acknowledges support from the Research Council of Finland project 350458.
The Gravitational-wave Optical Transient Observer (GOTO) project acknowledges the support of the Monash-Warwick Alliance; University of Warwick; Monash University; University of Sheffield; University of Leicester; Armagh Observatory \& Planetarium; the National Astronomical Research Institute of Thailand (NARIT); Instituto de Astrofísica de Canarias (IAC); University of Portsmouth; University of Turku. We acknowledge support from the Science and Technology Facilities Council (STFC, grant numbers ST/T007184/1, ST/T003103/1, ST/T000406/1, ST/X001121/1 and ST/Z000165/1).

This publication uses data generated via the \url{Zooniverse.org} platform, development of which is funded by generous support, including a Global Impact Award from Google, and by a grant from the Alfred P. Sloan Foundation.
This research has made use of data and/or services provided by the International Astronomical Union's Minor Planet Center.
\subsection*{Software}
This research has made use of \textsc{astropy} \citep{astropy2013,astropy2018,astropy2022}, \textsc{geopandas} \citep{Jordahl2020}, \textsc{iraf} \citep{Tody1986,Tody1993}, \textsc{matplotlib} \citep{Hunter2007}, \textsc{numpy} \citep{Harris2020}, \textsc{pandas} \citep{McKinney2010} and \textsc{scipy} \citep{Virtanen2020}.
\section*{Data Availability}
GOTO images and source catalogs will be made available in a GOTO data release at a later date. Anonymised and/or aggregated classification data are made available upon reasonable request to the authors, but are anticipated to be released publicly at a later date. User-level Zooniverse data and PII will remain private following the Zooniverse User Agreement and Privacy Policy --\url{https://www.zooniverse.org/privacy}.



\bibliographystyle{mnras}
\bibliography{references} 



\appendix
\onecolumn
\section{Full List of Volunteers} \label{all_volunteers}

We are truly grateful for the extensive effort of our volunteer scientists in making the \textit{Kilonova Seekers} project happen. A full list of names of contributors (who gave permission for their name to be shared) since our launch is given below in alphabetical order, correct as of time of manuscript preparation:

\noindent
\footnotesize
5, 958bacsal, A, A Piras, A Taylor, A\_lot\_of\_imagination, Aaboli Samant, Aarush Naskar, Abby, Abditory, Abdulla G. Asanar, Abdurahman Mohamed, Abel, Abhijeeth Veeranki, AbrilPerezH, abrosio, achmadsujana, Ada Ji, Adam, Adam Cash, Adam Gibson, Adam Martinez, Adam Schufeldt, Adam Straub, adamzwawy, Adekunle Adejokun, Aditi Brij, Adrian Morales, Adrian Smith, Adrianna Jones, Adrien Droguet, Afjal\_khh, Agnid Nandi, Aguirre, Ahmad azizisani, Ahmed Estiak, Aiden Chadwick, Aimee Gonzalez Ferreira Sirvani Valentim, AJinSA, Aki Suvanto, Akiko Inamoto, aknepeter, Al Lamperti, Alaa Salah Afifi, Alan Teague, Albert, Albiona Leka, Alejandro, Alejandro Arróliga Vanegas, Alejandro Lopez, Aleksandr, Aleksandr Ketov, Aleksandr Timofeev, Aleksandra Pogorzelska, Alex, Alex Al-Sammarraie, Alex Andersson, Alex Gabriel, Alex Lammers, Alex Mitchell, Alex Zuniga, Alexander Becker, Alexander Blagrave, Alexander Davidson, Alexander Doens, Alexander G. Plasser, Alexandra Hercilia Pereira Silva, Alexandre Celier, Alexia Fotini Panagopoulos, Alexis Carrilllo, Alexis Casey, Alexis Daniel Gómez Alatorre, Alexis MANET, Alexis Tombrello, Alfredo Gimeno, Ali Kiwan, Ali Reza fani, Ali Tejani, Alice, Alice Bull, Alice Hu, Alima, alimamo, Alina Borissenko, Aliona Philippova, Alison Edwards, Allison Myers, Allison Umberson, Alma, almalthea, Alvin Echeverria, Alyssa Chandler, Amadeus Gabriel dos Santos Siqueira Silva, Amanda, AMAR PAL SINGH, Amaury Vincent, Amber Alvidrez, Amelia Chaber, Amirali Shahriarymanesh, Ammar Vora, Amoli Kakkar, Amy, Ana, Ana Haag, Ana Karen Tapia, ANA LUIZA MAXIMO AGUIAR DE ALMEIDA, Ana M. Pizarro Galán, Ana Paula Waaijenberg, Ana Sofia de Oliveira Caldeira, analemma.sky, Anamaria Liana Axinte, Anargha Bose, Anastasia Eriksen, Anastasia Prybytko, anastasia scoggins, Anay Mishra, Andrea Bortoluzzi, Andrea Espinoza, Andrea Nava, Andrea Serio, Andrea Williams, Andrej Coleman, Andres Eloy Martinez Rojas, Andrew, Andrew Bickley, Andrew Boyer, Andrew Conan, Andrew Cooper, Andrew Del Santo, Andrew Obara, Andrew Shaw BSc(Hons) MCPara MRi, Andrew Waldie, Andrew Winkelman, Andrey Korobkov, Andrii Dzygunenko, Andry Nasief, Andrzej Bobinski, Andrzej Wojtowicz, Andy Tonthat, AndyTheAstronomer, Anel Madrigal Gonzalez, Angad Chadha, Angel Elbaz Sanz, Angela Brito, Angela Volpe, Angelika Reithmayer, Angelique Reder, Angelo De Lemos, Anil Vasudev, anita martins da cruz, Anita Springer, Anna, Anna Andriyanov, Anna Batueva, Anna Brisa Micheff Soares, Anna Clara de Souza Fraga, Anna Kruchinina, Anna Mackiewicz, Anna Plum, Anna Scott, Anna Vorobeva, Anna Zanone, AnnaJewel Pace, annparker, Anond Disyatat, anthony, Anthony R. Wells, Anthony Rainone, Anthony TREMBLIN, Antonio, ANTONIO JEFFERSON MONTE ALVERNE PAULINO, Antonio M. Puertas, Antonio Pasqua, Antony Davi Costa de Sena, anwilk, Anylem Gonzalez, Anđela Mogin, Aoife Boyd, Aoiffe Boyle, Aparna Joshi, Archana, Ariana montes, Arianne Ambion, arianny caetano, Arkanar, Arkaprova Dutta, Arkhipova Daria, Arla Heikkinen, Arlind.S, Arman Svoboda, armandina gutierrez, Armando I Zamora, armydragon637, Arnaud Dufourcq Lagelouse, arsama, Artemii Krykun, Arthur Almeida, Arthur Meunier, Arthur P. Pereira, arthur pereira martins, Arttu Sainio, arturovasquez, Artyom Yakubov, Aryan Vinod, Ash Washburn, Ashlee Kephart, Ashleigh Goh, Ashley Abrego, AShley Wilkinson, Ashley Willis, Ashton, Ashtyn Gibbs, Ashutosh, Ashwin Shenoy, Asim, asterisk\_man, Athanasia Vlachou, Atlas, Aubrey Tyson, Aurelijus A. Alekserius, Auriam, Aurora, Auryne, Aurélien GENIN, Austin Hughes, Axavier neyra, Axel Geovanni, Aya Ahmad, Aydın AYBAR, Ayushmaan Mishra, B L Goodwin, badgerfish, Baiba Dislere, Barbaa, barbara england, Barbara Hartmann, barmet76, Barrie Matthews, Bartlomiej Krajewski, Basar Anil, Basil, Basudev Bhattacharya, Basundhara Maji, Bawan Aziz Muhemed, bdinti, Beatriz Barros Maia, Beau, bekind2all, Bella Karlisch, Ben Bartel, Ben Cole, Ben Kelahlyah, Benjamin Kapsch, Benjamin Olson, Benjamin Pumphrey, benjamin savageau, Benjamin Zahradnik, Benoit ROUSSEAU, Bent Løschenkohl, Bernd Nikolaus, Bernhard, Bernice Buan, besharp, beta\_cigni, Beth Meeker, BHARAT GUPTE, Bhavesh Sai Arambakam Madhu, Bianca, Björn Wilde, Blaize Baehrens, Bob, Bob Birket, Bobbi Marcum, Bogosi Sekhukhuni, Bokre Samson, BorisBanjac, Boundlessness, Braden Hancock, Brady Lundin, Braiden king, bramboro, Brandi Halloran, Brandie Nuckolls, Brandon Adcock, Brendan, Brennen Boyer, Brent O'Connor, Brett Reilly, Brian Andersen, Brian cloke, Brian Nevins, Brian Spirk, Briana Gulas, Brianna, Bridget Foster, brinlong, Brittany Brockenton, Brix Ola, Broc Daly, broe317, Bronwyn Wallworth, Bruce Griego, Bruce Horlyck, Bryan F. Smith, Buldris, buzzwon, Byron allen begley, C Unsworth, C. D'silva, C. Luke Gurbin, C. S. Tolliver, Caballero, Gabriel D., Cairo Taylor, Calvin D Nourse, Cameron Alexander, Cameron Johnson, Cameron Lopes, Camille Mumm, Candela, CANNIZZARO, Carl Setzer, Carla V. Mejia, carloartemi, Carlos Alfredo Narváez Gaitán, Carlos Antonio santos, Carlos Augusto Araújo Silva, Carlos Nunez, Carmen Mandel, Carol A. Schneier, Carol G Taylor, Carolina Bresciani, Carolina Dos Santos Casaleiro Da Silva, Carolyn Bolus, Carolyn Sill, Carolyne Brough, Carrie black, Carsten Meldgaard, Carston Rose, Carter Hathaway, Caryme Martinez, Casandra Martin, Casey Bonham, Cassie Merkel, Cath Cockeram, Cath Sharp, cathcollins, catherinebp, Cauã Filipe Ribeiro Albuquerque Silva, Cecilia Lomax, Ceilidh Macrae Kirk, Ceona E., Cezary Kruszewski, Chan Hwee Im, Chappers34, charbel saliby, Charles Pennison, Charlie Frost, Charlotte Williams, Chase, Chasity Newland, Chayse Jones, chemistinside, Chen Shaojie, Chen Stanilovsky, Cherridah Weigel, Cherrine Wilder, chhanda bewtra m.d., Chiara palmitesta, Chinabob, ChipFaust, Chiroko, Chloe Ernspiker, Chloe Greenbaum, Chloe Le Lacheur, Chloe McElroy, Chris, Chris Barbosa, Chris McDaniel, Chris McFarlane, Chris Mitchell, Chris Nowlan, Chris Pattison, Chris Theofel, Chris\_bushell, chriscasper, chrisfro, Christian, Christian Sergienko, christine groen, Christine Lee, Christopher B. Davis, II, Christopher Bowen, Christopher Horga, Christopher Pemberton, Christopher Strauss, Christy Browne, Chuck Henrich, Cian Maestri, Ciaron Drain, Ciro Sirio Perrella, Claire R. Hadley, Claire Volinski, Claude Cornen, Claudia Gonzalez Lozano, Claudio Correa, Cledison Marcos da Silva, Cliff Kurlander, Clifford Brown, Clément Violette, Cody Cook, colcol, Cole, Cole Murphy, Colin Chandler, colin\_hewitt, comface, Connor Sands, Cooper Evans, Cooper Kelly, Corey McInerney, Cory Chambers, Craig Foss Olsen, Cristiano Secci, Cristina Almeida, Cristopher Cojocaru, cs192, csprucefield, CThomas, cubby348, curlytoplu, cwilton, Cynde, Cynthia Jerez-Lema, Cynthia Moore, Cyril, Cyrus Trial, Céline de Ruiter, D Brough, D J Spruce, d.gordon.banks, d\_ashenden, dadotron, Dale sinclair, Dalia Garcia, Damian Gleis, Damian Janson, Damien Jackson, Damien Laouteouet, Dan Ryczanowski, Dana Lubow, Daniel, Daniel Alquizaleth, Daniel Amaya, Daniel Berliner, Daniel Conte, Daniel Gadomski, Daniel Henley, Daniel J. Reisner, Daniel Karnuakh, Daniel Leibman, Daniel mireles do nascimento, Daniel Raso, Daniel Wolf, Daniela Gallego Ramírez, Danielle Perkins, Dannis Vo, Danny Cameron, Danny Campbell, Danny Roylance, Danveer Kalliecharan, Daria Machina, Darien, Darien Lefort, Darius Gumuliauskas, DarkAryan, dash\_5, Dave Anderson, Dave D, davews333, Davi Cordeiro dos Santos, Davi Lima Alcântara, David Akhmadullin, David Baker, David Briggs, David John Flood, David López Martínez, David Meierhenry, David R Harris, David Saewert, David Stefaniak, Davide, Davide Iannone, DavidFoss, davidselfe, Dawn Sturgeon, dcortesi, Dean Santos, deanroberts, DEBAYUDH CHAKRABORTY, Deborah Kelsey, Deborah Woods, Declan Raven, Deen, DEEPAK, Deirdre Harris, deivad, Dena A Mitchell, Denilso G. Delfrate, Denis, Denis Hathaway, Denis Pilon, Dennis Rowland, Dennis Toy Jr, Derrick Wales, Destin Smith, Deviek, Devon Gerik, Devrit Saha, Dhruv visariya, Dhruvatara Bhogishetty, Dhuertas, Diana Sironi, Didac Invernon Campoy, Diedre barnett Garcia, Diego Diaz, Diganta Sonowal, Dimitri, Dimitri Ferreira Lima, Dimitris Mitsikaris, dirkie, dj\_tjitso, Dmitriy, Dmitriy Korovin, Dmitrujs, Docwill8, dom\_mercer, Dominik Siefert, Dominik Swiniarski, Dominik Valouch, Dominika, Don, Don Feldman, Douglas Higgs, Douglas Madzier, Dphr, Dr Peter Musk, Dr Sabrina Gärtner, Dr. Brian Decker, Dr.T.K.Subramaniam, Drew, DrKlahn, drokly, Duangruetai Samransanit, Dubravko Jakovljevic, Duncan Grant, DUPONT Florent, Dylan Drazek, Dylan Jusino, Dylan N. Weinrich, Dylan Owen Reserva Unas, Dávid Fülöp, E Pratt, E. Mayr, E.N.G., Eaden Morton, Ebubekir Sark, Edgar Guzman-Contreras, Edgard Schwarz, Edna Soto, Eduarda V Baldo, Edward, Edward Caplin, Edward Mokurai Cherlin, Edwardo Garcia, Ege Turker, Ekin Alp Arslan, elandale, Eldhie Joy Rosales, ElectraVentures, Eleftheria Travlou, Elena Akimova, Elif Bayat, Elisa Di Dio, Elisabeth Baeten, Elisabeth Brann, elisah, elissa steele, Eliyah Palamarchuk, Elizabeth, Elizabeth Gall, elizabeth serna, Elizabeth Swope, Elizaveta Svitova, Ella, Ella Katkova, Ellie Gold, Elliot Jones, Elyssa Smith, Emanuel Agapios, Emi, Emilia Domingos, Emilie Wuattier, Emils Locmelis-Lunovs, Emily Burrage, Emily Jayne Bean, Emma Boyett, Emma Fagan, Emma Ryan, Emma Sarkissian, Emmett Hein, emptylica, epv95ngc, equidad1, erez dagan, Eric Bellm, ERIC FABRIGAT, Eric Kim, ERIC MAILLOT, Eric Peuster, Eric Yachen Zou, Erica J Welborn, erichill, ericjpaquin, Erick Gomez Lopez, Erik Rodriguez, Erin Brache, Erin Comparri, Erin Norris, Erin Zorzy, Ernest Jude P. Tiu, Ernst Schneidereit, Esmeralda Gonzalez, Estelle Baude, Esther Liufu, ET\_Junior, Ethan, Ethan Alday, Ethan Atkinson, Ethan Estey, Ethan J. Keefe, Ethan Vice, Ettore Fernandes Damique Aguiar, Eugene Mercado, eugenius, Evan Barber, Evana Shrestha, Evangelos Batzios, Eve, Evgeny Epifanov, evyn, Ewout Kerklaan, ExavierMcLeod, expofever, Eyob, Eyvindr Leavenworth, Ezequiel Santos Couto, ezflyer, fabienmazieres, Fabián Bacca Alvarado, Fabrice Lamareille, Fabrício Fachini, Felicia Yllenius, Felipe Laruccia Sant Anna, Felipe ranzani de Luca, Femke de Vroome, fierybrunettlass, filippp, Finley Saville-Brown, Finn Suratt, Fiona Ellis, Fiona McNeill, Flaviano Santos dos Reis, fleuger, Floor Goossens, flya200, FOURNAISE Alexandra, Franchesca Flowers, Francis Varley, Francisco Alexander Balmaceda VII, Francisco Zala Rucabado, Francois DUFOURMANTELLE, Frank Decapio, Frank Helk, Frank Stuart, Fred, Fred Hellmig, Freddie Hason, Frederic Elcin-Coolidge, Freya, frozenchosen, Fujai Muhammad Charieth, futterwacken, Fuyuki Remix, FZolee, G Castro, G.W, Gabriel Jaimes Illanes, Gabriel Lawrence, Gabriel Palacios, Gabriel Stewart, Gabriella Costa de Souza, Gabrielle Mendonça, Gamar Alsadah, Ganymede3, Garrett Cornwell, Garrett Smith, Gast\'on Gonz\'alez Kriegel, Gaudin Titouan, Gautham Arun, Gavin Dukowitz, Gavin SLoan, Geert Dankers, Gemini Smith, GeminiNoSaga, Geof Wyght, Geoff Keeling, George Bowers, George G. Guilkey, George Humberstone, George Kokaev, George Luker, Georgia Lock, Georgina Fernández Belmonte, Gerald W. Nash Jr., Geraldine Qiu, Gerard Planelles Ripoll, Gerrit Bischoff, Gert Jan Klootwijk, gfox, Gianluximon, Gianni Tornaghi, Giovanni, Giovanni Aparicio, Giovanni Colombo, Giselle Sanchez, Giulio T. Forcolin, Giuseppe Conzo, gjcolburn86, Gloria (preferer  George please), Gloria Hernandez, Glorii, goggog, Golden Wolter, Goowithabrain, Gord Harmer, Gorka, Gorobets Dmitrii Andreewitch, Grace Mere-ana Ashby, Grace Parker, Grace Waller, Grace Wells, Graeme Bartlett, Graham Parlett, Grant Larsen, Grantham Norris, greenfield05, Greg, Greg Borders, Greg Gajer, Greg Schwitzer, Greg Scott, Gregg Kerlin, Gregory Aydt, Gregory Lewis, Gribol, grosbeak, gryphachu, Guillermo S\'anchez Calvo, Guoyou Sun (\begin{CJK*}{UTF8}{gbsn}孙国佑\end{CJK*}), Gurmanavdeep Singh Mahal, Gustavo Afonso Gomes, Gustavo Manzanilla, Gwendolyn Cardente, gwhw, Haaniya Khan, Hakkı Alp Tekin, Haley Smith, Hali Edmunds, Halley Solanum Theia Janus Culver, Halvor Nafstad, Hangar77, Hannah DiBenedetto, Hannah Foltz, Hannah Martin, HarpiaLC, Harriet Tyler, Harry Adams, harsh mahajan, Harsh vardhan, Harshdeep Singh, Hasan Arda Güler, Hatim Piplodwala, Haven Tyler, hawkman, Heather Ritter, Hector M Castro, Heidi, Heidi deVeyra, Helen Bates, Helen Spiers, Helena Jane Gomez, hellkr, Heloísa Pascoal de Souza, Henning von Hoersten, Henry Gagnier, Henry Rauch, Henryk Krawczyk, Hernán Flecha Alfaro, HerrStahl, hiba farrukh, Hiba Mohiuddin, highwaystar, Higor Gabriel jadjiski soares, hiko, Hilary Johnson, HippyPhysicist, Hiruve Gallo, Hisato Hayashi, Holen Yee, Hristo Delev, Hugo Andrés Durantini Luca, HummDinger, Hunter Burke, Hushaan, HypnotiQ, Ian Banbury, Ian Barber, Ian Branigan, Ian Chu, Ian Kennedy, Ian Lin, IanH84, Igor Akeliev, Igor Korotskin, Igor Kuchik, Igor luiz lein martins, Iliana, Iliq zlatanov, Illyana Weinzetl, Ilyas Wajahat Zafar Jalisi, imdra, Ine Theunissen, inge janson, Inken Gatermann, Irina Thome, Isaac Wardell, IsaacPerks, Isabella Read, Isabella Suzanne Valentine, Isabelle\_bourgeois, Isabelli do Vale Silva, Isac Oliveira Leite, Isadora Velloso, ishaan kolipaka, Ishita Jaisia, Isidora martínez, Issy Walker, Istiu, Isza Denise De Jesus, Ivan Martin, Ivan Titov, Izabel Bramlett, J N, J. Furst, J. J. Dziak, J. Oliveto, J. Toth, Jaana Kemppainen, Jacek Jackiewicz, Jack Anderson, Jack R. Brelsford, Jackarific, Jackson Tomaszewski, Jacob Balch, Jacob Hanini, Jacob Rogers, Jacob Schmidt, Jacob Thadius Giggey, Jacob Williams, Jacqui S, jacquiejh, Jade Friedlis, jadkinssd, Jagadeesh Pitchai Pazham, Jahcari, Jaime Frankle, Jake Chon, Jakub Kowalik, James, James Galla, James Garland, James Goerke, James H Kinsman, James Hewitt, James Pearson, James Smith, James Wilson, jamicze, Jamie Bjune, Jamie child, Jamie Griggs, Jamie Ramsay, Jamie Thompson, Jamie Wyman, Jamon, Jamy547, Jan Jungmann, Jan Slavický, jan55, Jardin Nathan, Jari-Pekka Pääkkönen, jarphys, Jarwen\_, Jasmine Lao, Jason, Jason Daniels, Jason Griffith, Jason Singleton, Jatin Singh Tomar, javier, Javier Alvarez-Escalera, Javier Gonzalez Duran, Jay Darnell, Jayanta Ghosh, Jazz, Jbrabham, jddavidson, jean\_cool, jedkat, Jeff Hamner, Jeff Lesperance, Jeff Wilson, JEFFERSON LORENCONI DE MORAIS, Jeffrey Ruff, jelik, Jen Beck, Jennifer Burstein, Jennifer Kestell, Jennifer Krouse, Jennifer Penoyar, Jennifer Rackley, Jennifer Shearer, Jenny X. Zhao, Jeremiah Sisemore, Jeremy Maciolek, Jeremy Thomas, Jerico B. Azarcon, Jernalyn Dulza, Jeroen Pullens, Jeronimo, jess77, Jessica, Jessica Field, Jessica Shaffer, Jessica Vaccarino, Jesus Bible, Jesus Eduardo Ceron Sanchez, jgendera, Jhonatas Tokuno de Campos Firmino, Jian Sundvall, Jiashuo Zhang, Jillian Ropchan, Jim O'Donnell, Jim Paszternak, Jimena Bravo-Guerrero, Jimmy Fisher, jin young kim, Jingyuan Zhao (\begin{CJK*}{UTF8}{gbsn}赵经远\end{CJK*}), Jkmorse57, jlam21xp, jlynec, jmalnar, Joan Kalec, joanhopkins08, Joanna Jarmolowicz, Joanna Kaczmarczyk, Joanna Molenda-Żakowicz, Joao Pedro Santos, Joaquim Queiroz, Jocelyn Leon, Jodhviir Sekhon, Joe CC, Joe Lane, Joedube11, joeK2\_45, Johan Joby, John bowles, John C. Skorupski, John D. Krull, John Eltgroth, John Engler, John Falconer, John Gibson, John Haight, John Jossy, John Li Chen, John M. Cummins, John Martin Hunter, John R. O'Grady, John Simpson, John Welsh, John\_R\_Williams, Johnathan Gueltzau, Jolene Nethaway, Jon Bueno, Jon Nugent, Jon Paver, Jon Sutton, Jonah Donis, Jonah Gluck, Jonas Lehnberger, Jonas Nagel, Jonathan Hatton, Jonathan W. Landers, Jooheon Lee, Joonzoon, Jordan Newman, Jordi, jorge, Jorge A. Vilchez, Jose Alberto da Silva Campos, Jose Gabriel Nino Barreat, Jose Luis Perez III, Jose Reta, Josef, Joseph Brom, Joseph Constantine, Joseph M. Crisp, Joseph Molnar, Joseph Morrison, Joseph Vinik, Joseph-Michael Viggs, Josh Dean, Joshua, Joshua Adams, Joshua Green, Joshua Hottenstein, Joshua K. Sullivan, Joshua Malcolm garner, Joshua McArthur, Joshua Slauer, Joshua Tan, Joshua Thompson, Joshua Truong, joshwilde, Jovokna, Joxean Koret, Joyce Kimbrell, João Paulo Molina Moraes da Silva, jpvignes, JStarhunter, Juan Antonio Serrano, Juan Corti, Judith Kokesch, Judith König, Judith M. Kirshner, Judy, juhana, Jules van Horen, Julia Allison Urawski, Julia Augustin, Julia Ellers, Julia Hodges, Julian Van Allen, Julianne Register, Julien Cochet, Julien Ortega, Juliet Guttendorf, Julio César Evaristo Rosa, Junghoon Chung (Kyle), Justin Abramson, JustinPaulson, Juvenal Barry, Jyothsna Terli, Jürgen Saeftel, Kabir Singh, Kacper J. Kowalski, Kacper Zydron, Kaden Tro, kafter, Kai Macci, kainat, Kaitlyn R Deacon, Kanishka Faqiryar, Kara Alber, Karan Choudhary, Karen Babich, KarlPettit, Karolina Biskup, karu58, Kass Ulmer, Kat Elder, Kat Lakey, Katarína Kačicová, Kate Reddick, kateboyd, Katharine Harris, Katherine Alice Tylczak, Katherine Collins, Katie, Katie Simpson, Katja Novitskova, Katja Schilder-Leu, Katrina, kaufmann, Kay Williams, kay\_d, Kayla Briggs, Kaylee Coley, Kaylee Doll, Kayleigh Spriensma, Kaylie Scorza, KB Burson, KELLY BRADFORD KITCHENS, Kelly Lynne Bowell (Baulos), Kelly OLeary, Kelly Rolyns Dela Cruz, KellySun, Kelvin Haakmeester, Kendall, Kendall Dow, Kenedy Chauvette, kennedy velado, Kenny Cavanaugh, Kerry Knight, Kerttu Saatsi, Kerys Taryn Stevenson, Kesi Butler, Kevin Borrot, Kevin H. Smith, Kevin Healey, Kevin Kelaher, Kevin Lam, kevin reyes, Kevin Wailes, khamyn Collier Smith, KillerApp, Kim Berry, Kimberly, Kimberly Hovencamp, Kimberly Tangkion, Kimberly VanMoorlehem, Kimberlyn Ortiz, kimdonghyeon, Kings Cordova, Kiran Tikare, Kirby, Kiri L. K. Salazar, Kjell Nilsen-Nygaard, kkolodzie, kleistf, Korben, Korina Koci, Kostrikov Kirill, KOZlegend, krahling, kreole, kresimir.jednacak, Kris Hunt, Krissy Guttendorf, Krist Lim, Kristen Therese Roehrig, Ksenia Shutovska, ktarkin, Kyle Fitzgerald, Kylie Coakley, Kynko, Ladislav Vondrášek, lailawise, Lance Ganey, Lanny Huynh, Lara Arce, Larice Fields, Larry.Melanson, Lars Erik Kjellström, laughterfollowsme, Laura Brown, Laura Dionysius, Laura Trouille, Laurel Wolchok, Lauren Deinhardt, Lauren Gardner, Lauren Harris, Layssa Victória Barbalho Assunção Silva, lbibayoff, LDJ101, ldybug5012, Lea Frohring, Leah Callender, leelaht, Leila Stemler, leisanearthlink, Leka Sree Jaishankar, Lemartinel, Lena Blomeyer, Leonardo Alberto Colombo, Leonardo Rodrigues, Leonel Hégues Nava, Lezel Rabusa, Liam Ball, Liam Yanulis, Lidia A. Rendon Martinez, likethedeserts, Lincoln Hamilton, Linda Hoxie, Linda\_J.\_Berkel, lindemose, Linden Thompson, Lindsey Skinner, liondave, Liran Zhu, Lisa Cali, Lisa Rust, Lise Smith, Lissa, Liz Dowthwaite, Liz Erator, Lizbeth C Moya, Lizzeth Ruiz Arroyo, llviegas, Logan A. LaChance, Logan Buck, Logan Burlingame, Logan Fisk, Logan Sedgwick, Logan Warren, lolinda, Longueville Matthieu, lopsidedhead, Loredana Bellucci, lorenzlacson, Lorenzo Bay-Müller, Lori Romua, Louis, Louis A Masciocchi, Louis Sterobo, Louise St. Germain, Louise-Marie Bertrand, Low Koon Yen, Loïc Baert, Lrsafari, Luana alves antonio, Luana Mayara Rosa Afonso, Lubov, Luca Ubertini, Lucas, lucas, Lucas Marcoccia, Lucas Stroobants, Lucas Wheeler, Luci Jackson, Lucia Maria Ferreira Lima, Lucy chachka, LucyitSwD, Luigi Ferreira Lima, Luigi Lalonde, Luis Rosas, Luis Wieczorek, Luiz Eduardo Pugliese Ferreira, Luiz Fernando Teodoro Tabosa Flegler, Luke Hauser, Luke Holland, Luna Costa Antunes, luohua, LvdStraat, Lydia Watkins, Lynne Dale, Lynnette Colombo, Lyudmila Aleksyuk, M.S. Batukov, M.T. Mazzucato, Mable woodbury, Mackenzi Hill, MacKenzie Kaser, Madeleine Flowers, Madison Hall, Madisyn Shick, Magdalena Badura, Magdalini Christopoulou, Biologist, Magomed, Mahdi Zakir Mahi, Mainak Mondal, Maja Kuchcińska, Malcolm Fowles, Mallory Kelly, Manjit, manseincardiff, Manuel PONCET, Mara Clarissa Codog, Marat, Marc Hanset, Marc Philipp Riker, Marcela Moreira, Marcie Nightingale, Marcin Kobielusz, Marco Bortolazzi, Marco Tison, Marcus Camper, Marek Klčovanský, Marek Skalski, Margaret Walby, Margaret Zimmer, Maria, Maria Atef Thabet, Maria Clara Silva Bezerra, Maria Dubiel, Maria Eduarda de Souza Vasconcelos, Maria Elsa Rosales, Maria Fernanda Maciel, Maria Isabel Jarrin Nunes Nishiyama, Maria MADIK, Maria Martin, Maria Regina Ortiz, Maria Wicher, Maria-Olivia Torcea, Mariah t, mariakruglova, Marianne Hundling, Marie Tessier, Marika, marina alkiviades, Marina Corrêa Freitas, Mario Giovanola, Marisol Alvarado Resendiz, Marius Agafitei, Mark Abbott, Mark Hackney, Mark Hillaert, Mark mccormack, Mark Reynolds, Mark S. Hamblosa, Mark Stypczynski, Mark\_Moffatt, Markus Biburger, Markus Buchhorn, Marlon Braun, Maros Rolko, Marta, Marta Święch, martdoespenguinwatch, Martell Valencia Guerrero, Martha Casquette, martin hall, Martin Thomas, Marvin Thunderbolt, Mary Ann Brennan, Mary Falk, mary joy romano, Mary Kay, Maryangel Arenas, MaryV, MARZAK, María Cecilia Melgarejo Ferreyra, María Hernández, masch914, Massimo Bucklin, Masuma Akhter, Matalin Hansen, Mateusz Sałasiński, Mateusz Szałankiewicz, Mathias Steyaert, Mathieu Laroche, Mathilde Tempke, Matt Hill, Matt Naylor, Matteo Cianfaglione, Matteo Mosca, Matthew, Matthew Day-Lopes, Matthew Krause, Matthew Lawrence, Matthew Schlademan, Matthew Terren, Matthew Zimmer, Matthias Breimann, Matthias Kohl-Himmelseher, Matthäus Balderer, Maureen Roberts, Maurice A. Hippleheuser, Mauricio Pizarro Morgunovsky, max, Max Bhoyroo, Max Krievs, Max Read, Max Stephenson, Max Werner, Max Zakrzewski, MaXoN, May Suu Suu Myat, Maya, Maya Douglas, Maya Fiorentino, Maya lester, Mayahuel Torres Guerrero, Mayukha Dissanayake, mdlw, mdmcdermitt, Medina Fazlic, Meena Balakrishnan, Megan, Mehal, mehmet tükenmez, Mekki, Melanie Garcia, Melissa Arsaut, Melodie Angela Pangilinan, melody maddahian, Melyssa Danielle de Souza Pereira, Mercutio Cacérès, Merijn Benard, Mervyn Hing, mfwuk, Mg Kaung Sat Thant, Mia Burgener, mia hernandez, Micaiah Balonek, mich\_wimbledon, michael, Michael Antoni, Michael Borland, Michael D. Cochenet, Michael Demuth, Michael Douglas Soares Tobias, Michael Ellison, Michael Galbraith, Michael Hamblin, Michael J Morrin, Michael J. Adams, Michael J. Tolentino, Michael Langdon, michael mae, Michael Nunn, Michael Pearle Jr., Michael Phillip, Michael Please, Michael Stouffer, Michael Stutters, Michael W Heaven, Michael W. Zabarouskas, Michel Kluyskens, Michele Boldini, Micheline, Michelle, Michelle D. Espinoza, Michelle Gregory, Michelle Lucas, Michelle Moore, Michelle williams, Michkov, Miguel Angel Agüero Romero, Miguel González, Miguel Guilherme, Mihaela, Mihai Strömpl, Mika Sadikario, Mikayla Creitz, Mike, Mike Barrett, Mike Barton, Mike from Michigan, Mike Stirling, Mike Waldner, Mike Young, Mike4Tammy, Mila Mucha Jana Moleman, Pablo Moleman, Miles Johnstone, Miles Whatcott, Millicent Switzer, Millie Evers, Miss Fox James, Miss. Gunaretnam Keerthana, Missybee35, mitch, mitchell spiller, Miłosz, mmilson, mohamed kamesh, Mohammad Odeh, MOHAMMAD SAMEER, Mohammed Mahabu Subhani, Mohmad Sharif Jamali, Monica E., Monte J. VanDeusen, Morgan, Morgan Grimes, Morgan Jensen, Mori Joseph, Moses Fleischman, Mouad Derfoufi, Mr L, mrakes, Mreowwww, Mrityunjay Saxena, mroeling, msfletcher, mshimohi, MUHAMMAD AMIRUL AMIN B. ABD RAHIM, muhammadmustafa, Mustafa, Myar Misellati, Myranda Keightley, Márcia Saori Câmara Kishi, Márjorie Mourão, N. J. Leyland Elsom, Nadine Trimpin, Naomi K. Lee, naslund, Nastya Kulikova, Natalia, Natalia López Guzmán, Natalia Pirogova, Natalie Jones, Nataliia, Natasha A Frakes, Natasha Van Bemmel, NateA, Nathalie Amador, Nathalie Smits, Nathaly Hernandez, Nathan Coetzee, Nathan Fontanez, Nathan Scalzone, Nathan Yusuf, nathancbell, Navahra Lindsay, Navaneethakrishnan, Naveen kumar S, nchazarra, nebogipfel, Neil Cottrell, Neil Kenneth Burton, Neil Sharp, Neil Wickens, Nelvish Naiker, Nerine Storm Thompson, Nevaeh, NeZur, Ng Wen Xuan, Nica Roseme, Nicholas D. McGranahan, Nicholas Dunn, Nicholas Garrett, Nicholas Hans, Nicholas Hertz, Nicholas J Bianchi, Nicholas Phu, Nicholas Sheppard, Nicholas Yarbrough, Nichole Coryell, Nick, Nick CUmmings, Nick Weilert, Nickywan, Nicola Bird, Nicolas Agustín Menillo, Nicolas Cano, Nicolas Lamarche, Nicole Boisvert, Nicole Meireles Gomes, Nicole P. Vogt, Nicole Parks, Nicole Wolff, Niels Anonymous, Nigel Branson, nigel1960, nihar.mishra, Nik Ritchie, Nike Koper, Nikita Babin, Niko Strauch, Nikolaos Dimitrios Poulos, Nikolas Mair, Nikshey Yadav, Nisha, Nishant Deo, Noah Childers, Noam dos Santos Gomes Cerqueira, NoctornalE\_, Nolan Reket, NonLinearTurtle, Nova and Diva Thapa Magar, Noé Susol, Nymie Goat, Obverse, Olayeni Anifowose, Oleksandra Pyshna, Olena Kravhuk, Olesya Shaturova, Olga, Olga Stehle, Olga Stikhina, Oliver Linnhoff, Oliver Lykke Petersen, Oliver ROSSA, Olivia Bransford, Olivia Elgar, Olwenn Daniel, Om Rohith Karamala, Ondřej Koubík, Onni Tengström, Oren Buerke, Origami99, Ornella Guyet, OSCAR MEANI, Oscar Rivera, Oskar Blachut, Owen Pearman, Pablo Gonzalez sepulveda, Pablo Gómez Alcojor, pablo.rpoval, Paige Schedler, Paloma, Pamela Allen, pammins, Panthi Barot, Paramvir Kasana, Parker, Parker Betts, Parth bhawsar, Pascal Delcroix, Patrece Vreeswyk, Patricia Boyce, Patricia Fernandez, Patricia Kalimootoo, Patryk Krawczyński, Paul Fuentes, Paul Kirkland, Paula Sofia Paz Garcia, Paulo Gomes Monteiro, paulway, pavan, Payson Clausen, peadams1992, Pedro Berges, Pedro Garcia-Lario, Pedro Gonçalves, Pedro Guilherme de Barros Silva, Pedro Ricardo Balata dos Santos Costa, Peggy Hall, Pepito, Percival Nollido, PERNEY CLEMENTINE, Perry Lin, Pete Hermes, Peter Ansorge, Peter Barker, Peter Holmes, Peter Lange, Peter Langohr, Peter Wuestner, petersamuels, petestmarie, Petika, Peyton Sykes, philip.cross, Philipp Wenzler, Philippe Reclus, PhilipWang, Phillip Chen, Phoebe Lister, Phuong Duong, picerx38, Pietro Maiorana, pinegrove, Piotr Małek, Piotr Zając, Piotr Łojko, pippa brown, pitcher12k, pitchtwit, pjaj, planetari7, plent2d, Polina, Poovendhan Mathiyazhagan, PowlOwl, prachi, Prajwalita Jayadev Chavan, Pranav Gurusubramanian, Prasad Vengurlekar, Pratibha Rathore, Praza Kembaren, Prima\_Rosa, Princess Keila Nuñez, Pritam Dutta, Pritam Raikar, Priyansh Sharan, Prohor Thomas Polipartov, ProWang, Przemek, PsY4aLL, Punyae Bhatia, Péter Meglécz, Quaid Milholen, Quentin Laydecki, Quinn Torgerson, R. Hoornstra, Rachel A. Ford, Radoslaw Siewierski, Radosław Dobrzycki, Raella Rothman, Rafael Cifuentes, Rafaela Carolina Schmitt, Raghav Mishra, Rainey Waddell, Rajalekshmy, Ralph Allen, Ramesh Subramanian, Ramloo, Ramona Valková, Raphael, Rathika, ray van dycke, Rayen Dahmoul, readwin, rebecca Robinson, red mtn, RedTrev, Reem, RefugeZero, Reiko.Kawashima, Reina bonilla, Renee Appelman, Renyuqing, Rex\_Mundi, Rey Tolentino, rfsmith, Rhea Porter, Rhodri Thomas, Ricardo, Ricardo Costa Silva, Ricardo Meloney, Richard Bunche, Richard Cheal, Richard Edwardson, Richard Mukerji, Richard perry, Richard Ryan Rote, Richard Wambach, Rick de Koster, Ricky Grimaldi, Riku J. Laukka, Riley McManus, Rita Jónyer, Rivers Langston, rmhoek, Rob Ash, Rob Woodings, Robbie, Robert Apollo Pepper, Robert Auburger, Robert Deering, Robert G Lawless, Robert Jan Bauer, Robert Mazarelli, Robert Rice, Robert Stone, Roberta Reed, Roberto Haver, Robyn Wilson, Rochelle Adkins, rochelle thomas, rodnjes, Rodrigo De Souza janssen, Rogelio Calyecac, Roger Blake, Roger Feo, Roger Gagnon, Roiban Elena Monica, Rolf Agterhuis, Rolf Arndt, Ron Usry, Ronan Lucchesi, Ronnie Payton, Rory Campbell, Rory O'Connell, Rosemary Billington, Rosemary Patyus, Rosyid Haryadi, Rouanne Bohg, Rowan Callahan, Rowan Lavender, Royce Edward Sayers, Rubru, ruby, Rudra Protap Nandi, Rui Augusto Barbosa da Silva, RUMOR47, Rupal Maitra, Russell Fulton, Russell Klare, Ruth McCabe, Rutten, Ruud van Kruchten, Ryan, Ryan Hazen, Ryan J, Ryan Tansey, ryder, Rylie Gipson, Rylie Roche, Ryne Kreitz, Rönn Schück, S Ramya, S Uzel, Sabine Vollenhofer-Schrumpf, Sabnam Shrestha, Sabrina west, Sagarika Menon, Sage Alyxander, Sai Shankar, Saif Khwaja, Saint.Maria, Sallyann Chesson, Salvador, Salvador Araujo, Sam C., Sam Coletta, Sam L, Sam Rowcliffe, Sam Scarborough, Sam Swaim, Sam Watson, Samantha Anderson Mcauley, Samantha Cover, Samantha Crossfield, same, Sameh Faty, Samuel Anderson, Samuel Edward Lewis Halstead, Samuel hasley hunwick, Samuel Maragno Predolim, samuel rodrigues henrique, Samuel Williams, Sanchayan Sarkar, Sandra Lee Harris, MDiv, Sandro Montagner, Sanjith, Sankuratri, sanra, Saoirse Cox, saphira, Sara Gankhuyag, Sara Kindschus, Sara R, Sarah Grissett, Sarah Mae Raybould, Sarah R. Kleinman, Sarah Starks, Sarah Williams, Sarathkumar, Sari-Lynn Kerkhofs, Sasha Gdaniec, satsig, Satyajit Grover, Saumya Garg, Saurao M Upare, Sauro Gaudenzi, Savannah Gamache, Savannah Powell, sbfoulkes, scarab64, Scott C Evers, Scott Nantais, Scott Rogers, seadour, Sean Hohmann, Sean M. O'Brien, Sean Miller, sebastian abbatoy, Sebastián Alejandro Freigeiro, See Moua, Seema Shroff, Sema DEMİRBAĞ, Seojin Park, Seppo Hassinen, Sergey Sedov, Sergio, Sergio Roda, Serkan Alabaş, Serkan Kadi, Shadae Taylor, Shaik nissar hussain, Shakaraliyeva Leila, Shaleena Turner, shalnachywyt, shane Masterson, Shannon Adams, Shannon Kindilien, PhD, Sharad Pednekar, Sharon Fulton, Sharuyan, Shaul Lindblom, Shaun Coulon, Shawn Phelps, Shehab Khidr, Sheree Gillett, Shirley-Anne McIntyre, Shiv, Shivam Kumar Sharma, Shivam Saxena, Shivani Hancock, shocko61, Shogo Noma, Shotoya Simmons, Shree Ghosh, Shrey Ajaykumar Bhatt, Shreyas G Kamath, SHRIRAM, Shuming Wang, Sian Oleary, Siddharth, Siddharth Kuchimanchi, Siddharth Mehrotra, Sidhant Hargunani, Siegfried Zanier, Siham, simek, Simion Ochieng Agumba, Simon, Simon Barker, Simon Gardner, Simon Hennessey, Simon Lund Sig Bentzen, simon nicholson, simone marcorini, Simran Patel, Sinai Coons, Sivapriya Arvind, Skyboy777, skywarrior, slebedeva, smartalice, smith7748, snafu04, snappa, snikdekl, Sniwa, Sofia, Sofia G, Sofía Lorena Vázquez Montoya, Soham Thaker, Sohini Mukherjee, Somia Ditor, Somobrata Mitra, Sonny Black, Sonya Singh, Sophia, Sophia Ottner, Sophia Rehan, Sophia Rencher, Sophia Shannon, Sophie Whitley, Sovan acharya, Spencer Emmons, Srijan Banerjee, SRIJITA BISWAS, Sristi Betal, Srushti Pardhi, Stabitha Cru'ktfinger, Stacey E. Farr, Stacy Walker-Nez, starchitect, starsjosh, stasxh, Stephanie, Stephanie Colbert, Stephanie mosher, Stephanie Ochoa, Stephanie Santo, Stephany Palacios Martinez, Stephen Cole, Stephen Gazard, Stephen Kiraly, Stephen Molatlhegi, Stephen Patrick Brennan, Stephen R. Hyzer, steve, Steve Davies, Steve Lake, steve\_meadows, SteveF48, stevekrz, Steven Rosander, Stickywarp, Stitch1981, Stukalin Danil, Subhas Mazumdar, Sudeeksha Bhattacharyya, surhudm, Susanna Dargan, Susannah Bispham, Svetoslav Alexandrov, svr303, swaas, SY, Sydney Letlow, Syed Hozaifa Shabur, Syed Muhammad Abdullah Bin Hasan, Systrie, Tahseen Huda, Taira Agosto, tairving, Tan Poh Ching, Tania Ramirez, tania.ann.007, Tanishk Mohan, Tara, Tarant0ga, Tarpon Thompson, Tasfia Tarannum, Tatyana Reid, Taylor Benware, tdenton, tej Rajesh, tenfyr, Teodor Michalski, Terdoo Osu, Terrance P Felegie, Sr, Terry Georgas, Thaliyadath K Ravindranath, thecrazedlog, Theresa, Theresa Carl, Thomas BERTRAND, Thomas E Collett, Thomas Hoffmann, Thomas Jenkins, Thomas Kläger, Thomas Koceja, Thomas Laube, Thomas massey, Thomas R. Stewart, Thomas S Grady, Thomas Thomopoulos, thorsteinn, threepwood, Tia Rice, Tiep Tran, Tiffany Shaw-Diaz, Tim Dyson, Tim Jaramillo, Tim Loris Kunze, Tim Plank, Tim Pointing, Tim-Christopher Aust, tiorriot, tireniolu Oladipo, tkuhnle, Tobias Géron, tobidd, Tobrux, Toby Leeper, Tom, Tom Bickle, Tom Davison, Tom Osman, Tom Rich, Tomas Koval, Tomassci, tomjon, TomThaler, tony, Tony, Tony Grant, Tony Hoffman, Torsha Kundu, Torsha Majumder, tpetera, Trandafir Mircea, Trapeznikov Egor Valerevich, trash0, Travis Olah, Travis Rector, Trent Pierce, Tricia, trippyskippy, Trish Martinache, Tuesday Muse, TULSI GOYAL, Tyler Andreww, Tyler Bestram, Tyler Hirshberg, Tyronius McBasketball, Tyson Dabney, udayatejwani, Udunie, Uliana Shiliaeva, Uriel, Utsav Khan, Utsho, Vahavy, Vahid Kermani, Vaideeshwar Sivasubramanian, Vajrapani, Valentina, Valeria Tokareva, Valerie Blanco, valerie flores, Valerie Pegg, Valerie R seymour, Valmir Martins de Morais, Vandytim, Vanessa Eliseo, vanrock70, Varooni Manoj Sawant, Varsha, Vasileios Vlachakis, Vasiliy, Vasylenko A., Vaughan, venatrix, Venkatesh Deshpande, Vern Sowers, Viacheslav Zelenev, Victor, Victor Celli, Victor Cunha Da Silva, Victor Edwin, victor juan garcia porcel, Victoria Jackiw, Vikrant Kurmude, Viktor, Viktor Dobrenov, Viktor Wase, Vincent Hobeïka, Virgilio Gonano, Vishwanand Doobay, Vitor Luis Gonçales Dias, vivian, Vlad, Vlad Mihai, Vladimir, volare09, Walker Wells, Walter E Moody Jr, Walter MacDonald, wangqintao, warrenchen, Weatherly-Battaglia, WEBs\_in\_space, WELTON VAZ DE SOUZA, Wenceslao Santiago Germán, Wendy, Wendy Smith, wendy27, Wenjie Zhou, Wentao Huang, Weronika, Wesley Gabriel de Oliveira Melo, Wesley Teh Ee Wen, wesley webb, wgoltz, Wilfried Domainko, Wilfried Esken, Wilhelm A. Weidmann, Will, Will Haresch, William B Hernandez, william birney, William Midgley, William Paul Rhodes, William Russell, Willow Colson, Windsor S Smyser, WOJCIECH MIKINA, WolverineWazza, wrackard, Xander sprangers, Xavier Dartevel, Xenon Chase, xiaoguangliu, xKingx, YaBoy, Yanilsa H Marte Rodriguez, yanyam, Yavin\_4, YGBS, Yogurtz, Yohan Terpend, Yolly Reyes, Younes Oubkis, Yufan Fane Zhou, Yun Chen, Yuri Peruzzi, Yuri Sushkov, Yvonne Harrison, Zac Thomas, Zaccaria Vidali, Zach Ortega, Zach Schroeder, Zachary Fleisig, Zachary Thede, Zachary Vaillancourt, zacho, zak bennett, Zakck Goolsbee, zan, Zander Polk, Zarriah Fisher, Zdeněk Flanderka, Zebraorpanda, zedcat, zena barabandi, Zenith Diehl., Zhiyuan Cheng, Zijun He, Ziyun Tang, Zoe Kateri, Zoe Vickhouse, Zoryana, Zovacor, Ágata Moretti Zaneti, Íris Eduarda Forcel Roncada, \selectlanguage{russian}Александр, Владимир Асташин, Даниил, Дмитрий Дуда, Дмитрий Сергеевич Фёдоров, Евгений Юрченко, Матвей, Николь Вельковская \selectlanguage{english} , \begin{CJK*}{UTF8}{gbsn}杨皓添\end{CJK*}, \begin{CJK*}{UTF8}{gbsn}林于顺\end{CJK*}, \begin{CJK*}{UTF8}{mj}蔡頌恩\end{CJK*}, \begin{CJK*}{UTF8}{gbsn}马若瑜\end{CJK*}, \begin{CJK*}{UTF8}{mj}김경현\end{CJK*}, \begin{CJK*}{UTF8}{mj}이승우\end{CJK*}, \begin{CJK*}{UTF8}{mj}하리경\end{CJK*} 


\bsp	
\label{lastpage}
\end{document}